\documentclass[sigconf,nonacm]{acmart}
\usepackage{float, graphicx}
\usepackage{lipsum}
\usepackage{multirow}
\usepackage{tabularx}
\usepackage{csquotes}

\AtBeginDocument{%
  \providecommand\BibTeX{{%
    \normalfont B\kern-0.5em{\scshape i\kern-0.25em b}\kern-0.8em\TeX}}}
    
\makeatletter
\newcommand{\concat@}{\mathop{\vphantom{\sum}\mathpalette\concat@@\relax}}
\newcommand{\concat@@}[2]{%
  \vcenter{\hbox{%
    \sbox\z@{$#1\sum$}%
    \resizebox{\width}{\dimexpr\ht\z@+\dp\z@\relax}{\raisebox{\depth}{$\m@th#1\Vert$}}%
  }}%
}
\newcommand{\concat}{\DOTSB\concat@\slimits@}
\makeatother
\setcopyright{none}
\copyrightyear{2024}
\acmYear{2024}
\acmDOI{}
\acmConference[]{}{}

\begin{document}

\title{Integrating Representational Gestures into Automatically Generated Embodied Explanations and its Effects on Understanding and Interaction Quality}

\author{Amelie Sophie Robrecht}
\authornotemark[1]
\orcid{0000-0001-5622-8248}
\affiliation{
  \institution{Social Cognitive Systems Group}
  \institution{Bielefeld University}
  \streetaddress{Universitätsstraße 25}
  \country{Germany}
}

\author{Hendric Voss}
\authornote{Both authors contributed equally to this research.}
\orcid{0009-0003-3646-7702}
\affiliation{
  \institution{Social Cognitive Systems Group}
  \institution{Bielefeld University}
  \streetaddress{Universitätsstraße 25}
  \country{Germany}
}

\author{Lisa Gottschalk}
\orcid{0009-0007-8084-4439}
\affiliation{
  \institution{Social Cognitive Systems Group}
  \institution{Bielefeld University}
  \streetaddress{Universitätsstraße 25}
  \country{Germany}
}

\author{Stefan Kopp}
\orcid{0000-0002-4047-9277}
\affiliation{%
  \institution{Social Cognitive Systems Group}
  \institution{Bielefeld University}
  \streetaddress{Universitätsstraße 25}
  \country{Germany}
}
\pagestyle{plain}

\renewcommand{\shortauthors}{Robrecht and Voß, et al.}
\renewcommand\shorttitle{Integrating Representational Gestures into Automatically Generated Embodied Explanations}

\begin{abstract}
In human interaction, gestures serve various functions such as marking speech rhythm, highlighting key elements, and supplementing information. These gestures are also observed in explanatory contexts. However, the impact of gestures on explanations provided by virtual agents remains underexplored. A user study was carried out to investigate how different types of gestures influence perceived interaction quality and listener understanding. This study addresses the effect of gestures in explanation by developing an embodied virtual explainer integrating both beat gestures and iconic gestures to enhance its automatically generated verbal explanations. Our model combines beat gestures generated by a learned speech-driven synthesis module with manually captured iconic gestures, supporting the agent's verbal expressions about the board game Quarto! as an explanation scenario. 
Findings indicate that neither the use of iconic gestures alone nor their combination with beat gestures outperforms the baseline or beat-only conditions in terms of understanding. Nonetheless, compared to prior research, the embodied agent significantly enhances understanding.

\end{abstract}



\begin{teaserfigure}
  \includegraphics[width=\textwidth]{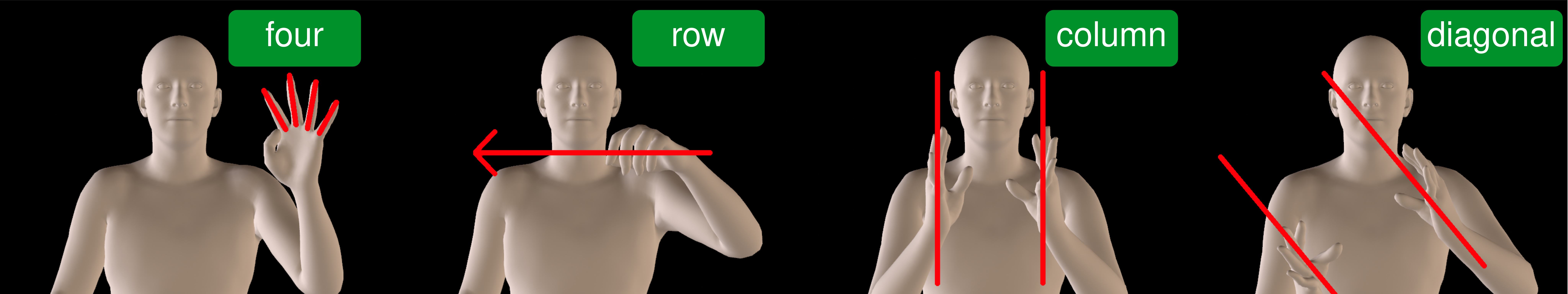}
  \caption{Examples of iconic gestures used by the virtual agent in the explanation of the board game \textit{Quarto!}}
  \label{fig:teaser}
\end{teaserfigure}

\maketitle

\section{Introduction}
Successful human communication involves multiple forms of modalities, including spoken language, facial cues, and body language. Understanding and generating these multimodal cues allows us to have meaningful and nuanced interactions in our everyday life \cite{cassell_speech-gesture_1999,wagner_gesture_2014}.
Current research on the social aspects of gestures is mainly focused on the effect gestures have on collaboration or interpretation tasks, with a clear focus on emerging gesture comprehension in children and young adults \cite{mcneill_iconic_1986, aussems_seeing_2019, zvaigzne_how_2019, magid_i_2017}. 

The importance of non-verbal communication for a successful collaborative or co-constructive explanation is long known \cite{mcneill_so_1985,lund_importance_2007}. Thus there is little research on how the explainee's understanding in an explanation is influenced by the performed gestures, especially with regard to studies that vary gesture parameters and quantitatively measure learning outcomes. As there is a lack of studies that vary gesture parameters, particularly on a fine scale, and quantitatively measure learning outcomes \cite{davis_impact_2018,sinatra_social_2021}, we provide such a study.
In addition, most insights into the effects gestures have on an explanation are from human-human interaction \cite{hostetter_when_2011} or focus on the human explainee expressing (mis)understanding \cite{lund_importance_2007}, while it is still unclear how far these effects can be transferred to human-agent explanations. 

Regarding the automatic generation of non-verbal behavior, the main focus on co-speech gesture generation currently lies in producing natural and human-like gestural motion from multiple input modalities \cite{nyatsanga_comprehensive_2023}. Although these gestures look natural, they still convey only very limited additional non-verbal meaning \cite{vos_augmented_2023}, and thus do not allow for the interaction quality achievable with virtual agents that employ iconic gestures in human-agent scenarios \cite{bergmann_individualized_2010}.

In this paper, we investigate the effects of different types of gestures on the objective understanding and the perceived interaction quality of a multimodal explanation. Our focus lies on studying the understanding and engagement of participants by distinguishing between shallow understanding and deep enabling \cite{buschmeier_forms_2023}. Therefore, understanding can be divided into comprehension -- the knowing that -- and enabledness -- the knowing how. Both forms of understanding can either appear in a shallow -- only surface knowledge -- or a deep -- being able to draw connections between information -- way.

To this end, we developed a multimodal virtual agent designed to explain the mechanics of the board game Quarto!. It is based on a novel model that complements the automatic generation of spoken explanations (from prior work) by generating four different kinds of gestures (baseline, beat, iconic, and mixed). An iconic gesture (illustrated in Figure \ref{fig:teaser}) conveys semantic information by presenting a depiction of the related aspects \cite{bergmann_increasing_2009}, while beat gestures are biphasic movements of the hand and do not carry any propositional content \cite{bosker_beat_2021}. In our current model, none of the generated gestures introduces new information to the explainee but instead augment the information already conveyed by speech. 
In the following, Sect.~\ref{related_w} provides relevant background about multimodality and gestures in human-human and human-agent interaction. Sect.~\ref{sec:themodel} presents the model for explanation generation, before Sect.~\ref{gestures} turns to the approach for beat and iconic gesture generation. Finally, a user study is presented (Sect. \ref{study_design}) and its results are discussed (Sect. \ref{sec:discussion}).

\section{Related Work}
\label{related_w}
While the benefit of gestures in human-human interaction is well known \cite{breckinridge_church_role_2007,wagner_gesture_2014}, the results in human-agent interaction are more ambiguous \cite{church_chapter_2017}. At present, little is known about the effect of the explainer's gestures on the explainee's understanding in human-agent explanation scenarios.

\subsection{Gestures in Human-Human Explanation}
An explanation can be seen as a co-constructive process, giving both interlocutors -- the explainer and the explainee -- an active role in the interaction \cite{rohlfing_explanation_2021}. As shown in \citet{lund_importance_2007} gestures in explanations are used for manifold reasons, they can refer to an object, end a verbal utterance, or accompany it to stress its importance. For human-human explanation, it has been shown that it helps the listener to understand the intended meaning and structure if the speaker uses gestures in an interaction \cite{breckinridge_church_role_2007, kita_what_2003}. It has also been shown that iconic gestures can support the long-term learning of second language vocabulary in children \cite{de_wit_effect_2018, bergmann_virtual_2013, belpaeme_guidelines_2018, rosenthal-von_der_putten_non-verbal_2020, rohlfing_how_2006}. 

It is a common approach to transfer human-human interaction patterns to human-agent interaction. For instance, there is research on how to transfer the explainer's understanding of an explainee's gestures from human-human to human-agent interaction \cite{rohlfing_how_2006}. Here, the authors introduce a theoretical approach on how agents can interpret posture and gesture based on behavior patterns in parent-child interaction.
So far, little is known about the effects of the explainer's gestures on the explainee's understanding and perception of an explanation in human-agent interaction. While there are implementations of gesturing agents for spatial description tasks \cite{bergmann_increasing_2009}, math equations \cite{perry_transitional_1988}, Piagetian conservation tasks \cite{ping_hands_2008}, or word learning in children \cite{de_wit_effect_2018}, to the best of our knowledge, there are currently no artificial explainers using specific gestures as information-conveying tools in their explanation.

\subsection{Gestures in Virtual Agents}\label{sec:gesturesinvirtualagents}

The generation and integration of nonverbal behavior in virtual agent scenarios have been a long-standing problem in virtual agent research \cite{kurokawa_gesture_1992, cassell_animated_1994}.
Early approaches mainly relied on rule-based models, with gesture templates created by hand, such as the Behavior Markup Language (BML) \cite{kopp_towards_2006,vilhjalmsson_behavior_2007} and the Behavior Expression Animation Toolkit (BEAT) \cite{cassell_beat_2004}, while more recent works primarily use deep-learning, graph-based or hybrid approaches to generate gestures from given input modalities \cite{liu_speech-based_2021, nyatsanga_comprehensive_2023, zhou_gesturemaster_2022, zhao_gesture_2023, vos_aq-gt_2023}. 

Regarding the effect of synthetic gestures on human-agent interaction, much work has been done on the perceived personality of the agent \cite{neff_evaluating_2010, liu_two_2016} or the creation of rapport with the agent \cite{gratch_creating_2007,bailenson_digital_2005}. It has been shown that users are more willing to engage in a human-like way if the agent uses human-like gestures \cite{kramer_effects_2007, parise_cooperating_1999}. At the same time, it has been shown that mismatching gestures have a measurable negative effect on such interactions \cite{salem_generation_2012, wagner_gesture_2014, kelly_neural_2004}.
In general, the main effects found are social perception effects (how is the agent perceived by the user?) and communicative effects (how does gesturing influence the course of interaction?). However, little effects have been found on user understanding and task performance \cite{church_chapter_2017, davis_impact_2018}. On the other hand, there are indications that participants benefit from gestures when it comes to learning, but there is not enough data to generalize from it \cite{macedonia_imitation_2014, davis_impact_2018}. 
Models based on co-construction face difficulty with repetition when users don't provide sufficient feedback, often due to underestimating the user's competence \cite{robrecht_study_2023}.
\subsection{Cognitive Load in Multimodal Interaction}\label{sec:cognitiveload}
There are different possible ways in which gestures can support or hamper the listener's processing and understanding of the utterance. We focus here on the cognitive load that gestures may impose on the listener, and which may particularly affect the processing of multimodal behavior of artificial agents. A lot of research has investigated the speaker's cognitive load and how it can be decreased by using different modalities \cite{oviatt_when_2024}. It has been shown that gestures help to structure an interaction and thereby minimize verbal load \cite{goldin-meadow_gesturing_2009}. \citet{chen_multimodal_2012} show which features are relevant to measure the current cognitive load and apply their method to different experimental scenarios. \citet{oviatt_when_2024} show that as cognitive load rises people tend to go multimodal, presumably to distribute the load over the used modalities. Only a few studies examined the impact of generated input on the listener. The question of how gestures affect the listener's cognitive load remains unanswered by them \cite{krieglstein_systematic_2022}.

 Gestures in interaction do influence both, the speaker and listener. Most research focuses on the positive effects gestures have on the person using them.
\citet{hostetter_comparing_2023} demonstrate that gesturing and other meaningful hand movements have a beneficial influence on verbal load.
 Similarly, research on memory calls showed that prohibiting the use of gestures diminishes the recall rate of memorized words \cite{matthews-saugstad_gesturing_2017}.  It has been shown that gestures also support the human explainer in structuring their explanation \cite{goldin-meadow_gesturing_2009}. When a non-existent object is described with gestures instead of speech, research has shown that the cognitive load reduces \cite{ping_gesturing_2010}. There is less research on how gestures are perceived, but studies show that presenting information on different modalities can expand the learner's capacity of working memory in learning scenarios \cite{mayer_split-attention_1998, tindall-ford_when_1997}. \textit{Cognitive load theory} states that presenting a task using different modalities (e.g. dual-mode presentation for solving geometry tasks) supports the expansion of working memory and helps to solve tasks \cite{mousavi_reducing_1995}.

In contrast to this, \textit{cognitive resource theory} shows that there can be a competition between modalities when performing a task, as they need to be processed in parallel \cite{wickens_compatibility_1983}. This approach also considers the cognitive load it takes to transfer input on one modality to a task that needs another modality.

\section{Explanation Model}
\label{sec:themodel}
The quality of an explanation is influenced by many aspects, such as adaptivity, multimodality, and information quality. As this paper describes the influence of gestures on understanding and perception of the explanation, all other aspects are kept as static as possible. We adopt a model for explanation generation, called SNAPE \cite{robrecht_snape_2023}. This model is capable of generating adaptive explanations, which have been shown to result in better understanding, especially better deep understanding, than a static explanation \cite{robrecht_study_2023}. This section will give a short overview of the model's approach and architecture.

SNAPE is based on a non-stationary Markov Decision Process (MDP) which evaluates the best action (which information to provide) and move (how to verbalize the information) dependent on the current internal model the agent has about the user. This model is called the Partner Model (PM) and consists of (1) an estimation of the user's current domain knowledge and (2) different global variables, such as expertise and attentiveness, which are based on the amount and quality of feedback that has been generated by the user so far. As transition probabilities and rewards are based on the PM, the MDP needs to be solved online using Monte Carlo Tree Search (MCTS). To keep the process real-time capable, the state space is kept small, which is grounded in the hierarchical structure of the model. A knowledge graph (KG) containing all necessary information for the explanation is extracted from an ontology, containing all possible information about the domain. Similar to the process observable in human-human explanations \cite{fisher_exploring_2023}, the KG is subdivided into semantic blocks. These blocks form the set of information that the MDP can use for inference. We extended the model by updating the template generation: the templates used for the verbalization of the explanation are now generated by a Large Language Model (LLM). This extension will be discussed in the following section\footnote{For further insights into the architecture, please refer to \cite{robrecht_snape_2023}, more information about complexity and preconditions are given in \cite{robrecht_study_2023}}.

\subsection{Using LLMs for utterance generation}
A key lesson learned from the previous model's performance, was that multiple pieces of information should be mergeable into one utterance if the current PM and the complexity of the information allow it. 
Hence, we have extended the SNAPE model to combine multiple triples into one utterance under certain conditions: (1) the triples have to be in the set of the five best next pieces of information to provide that is generated by MCTS, (2) they have to have the same linguistic move (provide new information, give additional information, repeat information, make a comparison), and (3) they have to share at least one entity. The vast number and complexity of potential triple combinations require a general and powerful approach to utterance generation. In the current version, all possible triple combinations are generated and matching utterances are created by prompting a fine-tuned Llama2 7B model \cite{touvron_llama_2023}. The model is fine-tuned on a dataset containing 487 items, each consisting of a list of triples, the move, and a matching output utterance. The fine-tuned model was used to generate multiple alternatives for each possible combination before running the explanation. The pre-generation allows to prevent hallucinations and minimizes the required computing power to keep the model real-time capable. 

\begin{table*}[]
\caption{Examples of Llama2 generated templates}
\begin{tabularx}{\textwidth}{lXX}
\toprule
\textbf{Move}               & \textbf{Triple}                                                   & \textbf{Template}                                                           \\
\midrule
\multirow{2}{*}{Provide}    & (Struktur, sein, Figurenmerkmal), (Groesse, sein, Figurenmerkmal) & Die Größe und Struktur sind Merkmale der Figuren.                           \\
                            & (structure, is, figure-feature), (size, is, figure-feature)       & \textit{Size and structure are features of the figure.}                     \\
                            &&\\
\multirow{2}{*}{Repeat}     & (Struktur, sein, Figurenmerkmal), (Groesse, sein, Figurenmerkmal) & Die Größe und die Struktur gehören zu den Figurenmerkmalen.                 \\
                            & (structure, is, figure-feature), (size, is, figure-feature)       & \textit{The size and structure are among the figure features.}              \\
                            &&\\
\multirow{2}{*}{Additional} & (Quarto, haben, Spielfiguren), (Spielfiguren, material, Holz)                                     & Die Spielfiguren bei Quarto sind aus Holz.                                  \\
                            & (quarto, has, figures), (figures, material, wood)                                         & \textit{The figures are made of wood.}                                      \\
                            &&\\
\multirow{2}{*}{Compare}    & (Spiel, haben, Ziel), (Reihe, sein, Ziel)                         & Das Bilden einer Reihe ist das Ziel von Quarto, genau wie bei Vier Gewinnt. \\
                            & (game, has, goal), (row, is, goal)                                & \textit{Forming a line is the aim of Quarto, just like in Best of Four.}  \\
                            
\bottomrule
\end{tabularx}
\label{tab:templates}
\end{table*}
As shown in Table \ref{tab:templates}, the prompts for template generation not only contain the information in the form of one or two triplets but also the linguistic move that the system is supposed to use to verbalize the information. If SNAPE introduces new information, the move is called \textbf{provide} information. For this move, information can only be taken from the knowledge graph. Providing information can either work or fail, which depends on the transition probability $T$ in the MDP. The transition probability is influenced by the currently inferred level of attentiveness the user has, as a user who gets distracted easily has a higher probability of missing information. If the move succeeds, the level of understanding $lou_i$ of this information increases. The strength of growth depends on the currently inferred level of expertise the user has. A \textbf{repetition} can be verbatim or reformulated, as long as no new information is given \cite{johnstone_repetition_1994}. Only necessary information can be repeated, accordingly the triple that is repeated has to be taken from the knowledge graph. Repetition is the simplest of the three available deepening moves. It has the highest probability of succeeding, but also the lowest increase in the level of understanding $lou_i$. The move \textbf{additional} adds information that is not necessary but potentially helpful to an already introduced, but not yet grounded, information. When considering giving additional information to the currently under discussion information $cud_i$, the model first needs to check for a fitting triple. A fitting triple is a triple that has to be part of the ontology but does not contain necessary information. Additionally, the triple needs to have the subject or object of the original triple as the subject. If a potential triple does exist, the move is an available action for the next step in the explanation process. A \textbf{comparison} gives supportive information to an already introduced triple. In this case, the triple is not taken from the Quarto! ontology, but from an ontology of another, comparable board game. Again, the triples have to share at least one entity, but can also be identical. Examples of each move can be seen in Tab.\ref{tab:templates}. 

\section{Gesture Generation}
\label{gestures}
In order to augment the explanations produced by the model described above with communicative gestures, we identified three main requirements for gestures to be used in an explanatory setting:
(1) The gestures had to be as human-like as possible and not distract from the given speech. (2) The generated gestures had to be extensible to incorporate additional iconic gestures in a natural and easily modifiable way.
(3) The gesture algorithm should be able to generate gestures in or near real-time, in a format that does not require extensive pre- or post-processing. 
To meet these requirements, we examined several different approaches to gesture generation, including but not limited to the models competing in the 2022 and 2023 GENEA challenges \cite{kucherenko2024evaluating, kucherenko_genea_2023}.
Since all diffusion-based implementations did not meet the real-time requirement and the GAN approaches did not produce satisfactory results in human evaluations, we focused on graph-based implementations. Inspired by the work of both \citet{zhou_gesturemaster_2022} and \citet{zhao_gesture_2023}, we developed a new graph-based gesture generation algorithm that enables the generation of realistic gestures in a real-time context.

\subsection{Dataset}
\label{gesture_dataset}
To create our new graph-based gesture generation algorithm, we first needed to get appropriate data that either already had iconic gestures or was usable in a base model on which iconic gestures could be added. Currently, iconic gesture data sets are almost non-existent. To our knowledge, only two annotated data sets for iconic gestures exist. The SaGA data corpus \cite{lucking_bielefeld_2010} with a small set of highly specific annotated data and the BEAT corpus \cite{liu_beat_2022} with acted interactions and rough categories for annotations. As we are mainly interested in natural interactions, we opted against the use of either of these corpora and instead captured our own data corpus. For this, we took 32 hours of TED and TEDx recordings with their subtitles \cite{ted_foundation_ted_2024, ted_foundation_tedx_2024} and tracked the body pose data for all recordings using One-Stage 3D Whole-Body Mesh Recovery with Component Aware Transformer (OSX) \cite{lin_one-stage_2023}. Afterward, we split the videos into individual video clips by detecting and cutting the videos along camera cuts using PySceneDetect \cite{castellano_home_2024}. Every clip was removed that did not include the main speaker, exhibited a low confidence rating during tracking, or had no perceivable movement. The final dataset consists of 24.2 hours of primarily beat gesture, text, and audio data. 

\subsection{Gesture Segmentation}
In contrast to deep learning-based approaches, which generate new gesture data from a trained multimodal data set, graph-based approaches are more closely related to retrieval-based techniques, where the input data is not used as training data, but as lightly processed chunking data in which the algorithm searches for an optimal path to generate new gestures \cite{zhao_gesture_2023, nyatsanga_comprehensive_2023}. Using audio, text, and gesture data as input, we first divided the entire training data into clips of length $s$ seconds, with an overlap of $\frac{s}{2}$ seconds. For our data, we chose a length of 2 for $s$. All clips shorter than $s$ seconds were discarded. We then performed individual data processing for each modality. For the text data, we encoded each word with fastText \cite{joulin_bag_2016,bojanowski_enriching_2017} and performed vector encoding for both the short historical context with $tx_1$ words and the long historical context with $tx_2$ words. We encoded both sequences using a MiniLM model \cite{wang_minilm_2020} and an openclip embedding model \cite{cherti_reproducible_2022, radford_learning_2021} trained on the DataComp1B dataset \cite{gadre_datacomp_2023}. As shown in equation $1$ and $2$, given the MiniLM model as $MiLM$, the OpenClip model as $Oclip$, the combination of the short and long historical sequence as $vr$, the combined sequence as $vt$, and the word at step $i$ as $w_i$, we formally perform

\begin{align}
vr_i &= \concat_{k=i}^{i-tx_1} w_i \oplus \concat_{k=i}^{i-tx_2} w_i \\
vt_i &= MiLM(vr_i) \oplus Oclip(vr_i)
\end{align} 
which concatenates the MiniLM an OpenClip vectors with the historic context $tx_1$ and $tx_2$.
For the audio processing, we resample the audio data to 24.000 Hz, and both compute a log normalized spectrogram with consecutive Fourier transformations on the raw audio data, as well as a vector embedding for the entire sequence $va$ using the wav2vec 2.0 model \cite{baevski_wav2vec_2020}. To more closely align the audio with the text data and to remove any superfluous information from the audio data, we additionally compute a processed version of the spectrogram and the wav2vec 2.0 vectors. For this, we compute a vector between zero and one, with a starting value of zero and the length of our sequence $s$, multiplied by the given frame rate of the original clip. We call this vector $ATs$. We iterate over every word in the sequence $Sx$ and note the start- and end timings $Ws$ and $We$, respectively. 
Using the start- and end timings, we calculate the middle point of the current word timing $Wm$. We linearly interpolate from the start point $Ws - 1$ with the value zero to the middle point $Wm$ with the value one, back down to the value zero at time point $We + 1$. If the word has a length of exactly one frame, we simply set this specific frame to one. Then, we multiply the vector $ATs$ with the spectrogram and wav2vec 2.0 vectors and save the resulting vectors.

For the gesture data, we kept the original position data unchanged. To achieve better coherence between the gesture generation segments, we calculated an additional vector $Gv$ by combining the first through sixth derivatives of the gesture position.

\subsection{Vector Embedding}
To align each dimension of the gesture, audio, and text data, we first normalized each dimension by subtracting the global mean and dividing it by the global standard deviation. As the gesture generation algorithm relies on matching the overlap of the referencing sequence (see chapter \ref{GG_do}), we split the data into two parts of length $\frac{s}{2}$ along the time axis and only kept the first half of the data.
As the given vectors exhibited a high dimensionality, we performed a Principal Components Analysis (PCA) on each data segment. We deliberately chose this reduction method instead of TSNE, UMAP, VAE, or PaCMAP dimensionality reduction, as all of these methods performed worse in terms of preserving the global structure of our data in our experiments \cite{maaten_visualizing_2008, mcinnes_umap_2020, kingma_auto-encoding_2022,wang_understanding_2021}. Using PCA, we reduced each dimension to a vector of length 256. Measuring the cumulative explained variance of the data, we preserved 92.6\%, 96.7\%, and 56.7\% of the variance for the gesture--, audio-, and text vectors, respectively. In order to enable fast nearest neighbor vector searches during the actual gesture generation, we stored the resulting vectors, split by modality and a combined vector of all modalities, in a vector database and computed a Hierarchical Navigable Small World graph (HNSW) over the entire vector sets \cite{group_postgresql_2024,kane_pgvectorpgvector_2024,malkov_efficient_2018}. Finally, we used the resulting vector database to create our vector graph, by defining every clip as a node in our gesture graph and performing an offline search for each singular modality and the combination of all modalities, storing the nearest 20 vectors as the edges for our graph.
Additionally, we added all edges to the graph that are naturally continuous and removed all edges that would lead to a reversed path (i.e. id:80 -> id:79) in the graph.

\subsection{Beat Gesture Generation}
\label{GG_do}
During the generation of new gestures, we split the given audio and text information into clips of length $s$, with an overlap of $\frac{s}{2}$ seconds. We calculated the vector data for the audio and text data for each segment, but contrary to the graph data, we only kept the second half of the vector data along the time axis. As the generation has no gesture information for the first sequence, we retrieved the gesture for the first sequence from our gesture database, by returning the result of the nearest neighbor search for the combined vector of audio and text. Starting with the second segment, we used the vector database to retrieve $Gn$ nodes, which minimized the L2 distance of our combined modality vector. For each given node, we calculated the best-performing path up to a depth of $Gd$, by traveling along the top $Ge$ edges for each node. Given the previous sequence vectors as $g$, the previous audio sequence as $a$, the previous text sequence as $t$, the current node vectors as $n$, the Euclidean distance as $d(x,y)$, and the result of the graph as $d_{step}$ we performed the following for every sequence step:

\begin{align}
d_{node}(x,y) &= \lambda_1 d(x_g,y_g) + \lambda_2 d(x_a,y_a) + \lambda_3 d(x_t,y_t) \\
d_{path}(vg_t,vn_t,t) &= \sum_{i=0}^{Gd} d_{node}(vg_{t+i},vn_{t+i}) \\
d_{step} &= \{d_{path}(vg_0,vn_0),...,d_{path}(vg_{Gn},vn_{Gn})\}
\label{distance_l2}
\end{align}

The hyperparameters $\lambda_1$, $\lambda_2$, and $\lambda_3$ were set to 4, 2, and 1, respectively. If a path contained an already selected node, the path was removed. After ordering the sequence list, we sampled from the top $k$ paths, by converting the normalized distances to a probability distribution. After choosing the sequence sample, we set the new gesture value to the gesture value of $Vg_0$ and repeated the process for the next step.

\subsection{Iconic Gestures}
\label{iconic_g}
As the main goal of the present work is to support the understanding of an explainee, we aimed to include believable representational (iconic) gestures that could be easily added and removed from the original gestures. To the best of our knowledge, no system exists that can automatically generate high-quality, aligned iconic gestures that are semantically coherent to a given verbal input. We hence manually annotated all instances of the generated explanation where an iconic gesture would make sense and pre-recorded suitable iconic gestures. For this, we captured 40 iconic gestures, with 3 different repetitions for variety, with a Logitech C920 in 1080p and used the same pose estimation algorithm as in Sect.~\ref{gesture_dataset} to extract the arm and hand positions \cite{lin_one-stage_2023}. As misalignment between gestures and speech is known to decrease interaction quality \cite{salem_generation_2012, wagner_gesture_2014, kelly_neural_2004}, we did not use any automatic method to determine the position and length of the iconic gesture clips, but instead placed, aligned, and blended all iconic gesture clips by hand using the software Blender \cite{community_blender_2018}. To ensure believable high-quality iconic gestures, we only added iconic gestures if new information during the explanation was given.

\section{Evaluation Study}
\label{study_design}

\begin{figure*}[!ht]
\minipage{0.5\textwidth}
  \includegraphics[width=\textwidth]{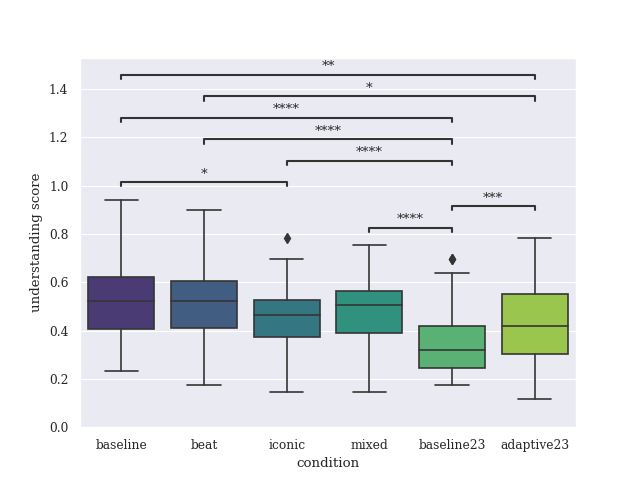}
  \caption{General understanding}\label{fig:general}
\endminipage
\minipage{0.5\textwidth}
  \includegraphics[width=\linewidth]{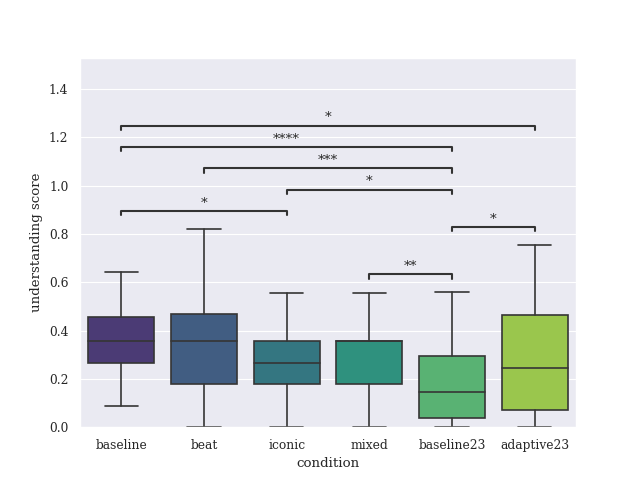}
  \caption{Deep understanding}\label{fig:deep}
\endminipage
\end{figure*}

We conducted an online user study\footnote{The study was preregistered in OSF: The anonymized pre-registration can be found at \url{https://osf.io/db7rz/?view_only=eddd075d791848599e009961b7352e68}}
to test the effects of gestures on the explainee's understanding and perception of the explanation generated by the virtual agent. The study has four conditions. In the baseline condition, the agent keeps the arms at its side and moves them slightly, but does not perform any gestures. In the beat condition, the agent performs the beat gestures generated by the gesture graph algorithm (Sect. \ref{gestures}). The iconic condition consists of the baseline condition to which we added the manually captured iconic gestures (cf.~Sect.~\ref{iconic_g}). Finally, the mixed condition combines the beat condition with the iconic gestures. \footnote{The videos for each condition can be found in the OSF project \url{https://osf.io/bf2yk/?view_only=704cf1241b4f4d7490bfb213a84c7fdc}}

Furthermore, we compare the user understanding with results obtained with purely textual but adaptive explanations generated by the SNAPE model \cite{robrecht_study_2023}. 
We perform a between-subject online study with 50 participants in each condition. We ensure an even split between male and female participants. The study language is German. Each participant is paid €12 / hour. After excluding outliers based on completion time, the analysis has a power between 0.79 and 0.8 for a Whitney-U-Test. The user perception of the interaction and the objective understanding are measured as dependent variables. The hypotheses we test are:
\begin{itemize}
    \item [H1] The user \textbf{perceives the interaction} as more positive in the mixed gesture condition.
    \item [H2] The user's \textbf{understanding} is higher in the mixed gesture condition.
    \item [H3] The \textbf{understanding} in the mixed gesture condition is higher than in the previous version of the agent.
\end{itemize}

\subsection{Results: Understanding}
The used understanding instrument is identical to the one used in \cite{robrecht_study_2023}
, allowing for a comparison between the SNAPE model and our multimodal explainer. The understanding instrument tests the effects on the task performance, which is one of the testable effects described in Section \ref{sec:gesturesinvirtualagents}. Before comparing with the adaptive agent, we compare the understanding of the different gesture conditions. The instrument contains two kinds of questions. The shallow questions test knowledge recall, in which the user has to decide whether a statement is true or false. The deep understanding test probes the user's ability to transfer the learned knowledge to in-game situations. Participants are asked to choose the best action in a given game situation shown as a picture. The questions are not only distinctive at the level of understanding depth, but also on forms of understanding \cite{buschmeier_forms_2023}. While the knowledge recall mainly tests the comprehension of the rules, the in-depth questionnaire tests to which extent the user can transfer and apply the learned knowledge: the user's enabledness. Thus, while most studies focus on the effects gestures have on long-term learning \cite{de_wit_effect_2018}, this study measures the immediate understanding the user has subsequent to the interaction.

Results show, that the understanding is not significantly higher in the mixed condition. The data is not normally distributed (Shapiro Wilk < 0.05 in each condition: general: p = 0.042, shallow: p = 3.094e-06, deep: p = 3.094e-06), so we performed a Mann-Whitney-U test instead of an ANOVA and posthoc t-tests. The Mann-Whitney-U test\footnote{ns: p <= 1.00e+00; *: 1.00e-02 < p <= 5.00e-02; **: 1.00e-03 < p <= 1.00e-02; ***: 1.00e-04 < p <= 1.00e-03; ****: p <= 1.00e-04} shows significant differences between the baseline and the iconic condition in the general (U = 1.302e+03, p = 3.310e-02) (Fig.\ref{fig:general}) as well as the deep understanding (U = 1.318e+03, p = 2.277e-02) (Fig.\ref{fig:deep}), while the difference is not significant between the conditions in the shallow understanding. To get a better insight into this effect, we compare the conditions for each question using the Mann-Whitney-U test. 
In the shallow understanding, only 3 out of 24 questions are influenced by the conditions. \textit{UN09 }(\textit{You can position any pieces on free spaces on the board.}, U=1.246e+03, p = 1.758e-02) and \textit{UN23} (\textit{Once you have finished a row, you have to shout ''Done''.}, U = 1.127e+03, p = 1.984e-02) show a significant difference between the beat and the iconic condition, while \textit{UN21} (\textit{To win, it is important to distribute the pieces randomly on the playing field at the beginning.}, U = 9.080e+02,  p = 3.248e-02) shows significant difference between iconic and mixed condition. In all three cases, iconic gestures produce significantly worse understanding. In contrast to the small number of significant items in the shallow understanding, 5 out of 8 test items in the deep understanding are significant: In \textit{UN31} (U = 1.231e+03, p = 2.958e-02) and \textit{UN32} (U = 1.262e+03, p = 2.774e-02) the baseline outperforms the iconic condition, in \textit{UN36} the beat condition produces higher deep understanding than the iconic (U = 1.203e+03, p = 4.183e-02) and the mixed condition (U =1.373e+03, p = 4.881e-03), while it only outperforms the mixed condition in \textit{UN37} (U = 1.225e+03, p = 4.504e-02). \textit{UN33} (U = 8.790e+02, p = 3.027e-02) is the only question, where the mixed condition outperforms the beat condition. When looking at the results for the individual deep understanding questions there is a trend for the baseline and beat condition to perform better than the iconic and the mixed condition\footnote{The plots for understanding of each question can be found in Appendix \ref{sec:appendixU}}. As shown above, this trend is significant for half of the questions. In sum, the results do not confirm the hypotheses but are rather contrary. Still, they show an interesting effect in this specific study: the understanding -- especially the deep enabledness -- is decreased by the usage of iconic gestures. Possible reasons for these results will be discussed in Section \ref{sec:discussion}.

Next to comparing the understanding of the different gesture conditions to each other, the understanding will now be compared to the understanding without an embodiment (\textit{baseline23}) and an adaptive explanation without an embodiment (\textit{adaptive23}). Although the explanations were given as text and the formulations were not identical to the formulations in this study, since they were generated by the SNAPE model \cite{robrecht_study_2023}, we consider this comparison meaningful as it allows to assess the overall explanation quality achieved by the embodied agent in this specific explanation domain.

When comparing the general understanding of the non-embodied agent to the updated version, all updated conditions do outperform the baseline23 (baseline: U = 2.666e+03 p = 4.096e-08, beat: U = 2.490e+03 p = 1.755e-06, iconic: U = 2.264e+03 p = 5.353e-05, mixed: U = 2.600e+03, p = 1.194e-06), but only the non-iconic conditions can outperform the adaptive condition (baseline: p = 6.304e-04  U = 1.843e+03, beat: p =2.078e-02 U = 2.214e+03).
When looking at the deep understanding, the embodied conditions also perform well in comparison to the non-embodied conditions. All of the embodied conditions generate a significantly better deep understanding (baseline: U = 2.479e+03 p = 7.813e-06, beat: U = 2.302e+03 p = 1.776e-04, iconic: U = 1.971e+03 p = 1.808e-02, mixed: U = 2.288e+03, p = 1.457e-03). However, only the baseline condition can outperform the adaptive condition when it comes to deep understanding (U = 2.275e+03 p = 1.640e-02).

\subsection{Results: Interaction Quality}
\begin{figure}[!b]
\minipage{0.25\textwidth}
  \includegraphics[width=\textwidth]{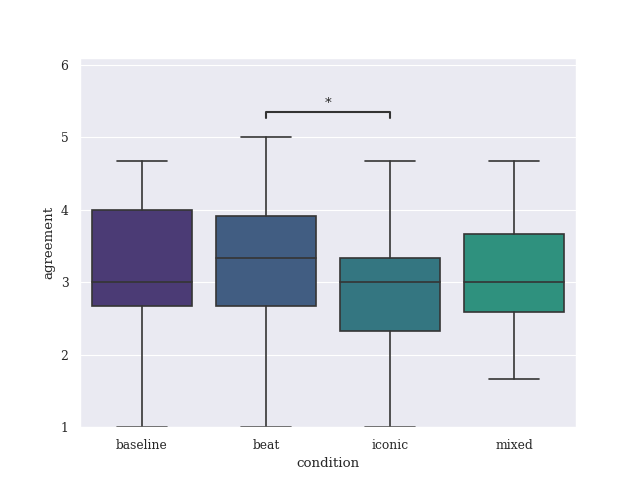}
  \caption{Trust}\label{fig:Trust}
\endminipage
\minipage{0.25\textwidth}
  \includegraphics[width=\linewidth]{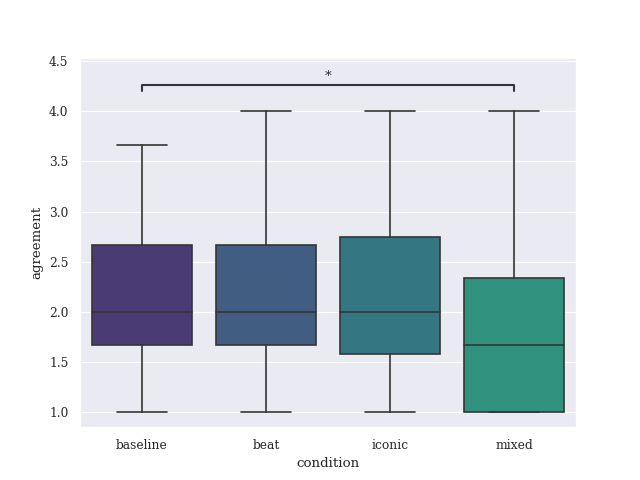}
  \caption{Likeability}\label{fig:likeability}
\endminipage
\newline
\minipage{0.25\textwidth}%
  \includegraphics[width=\linewidth]{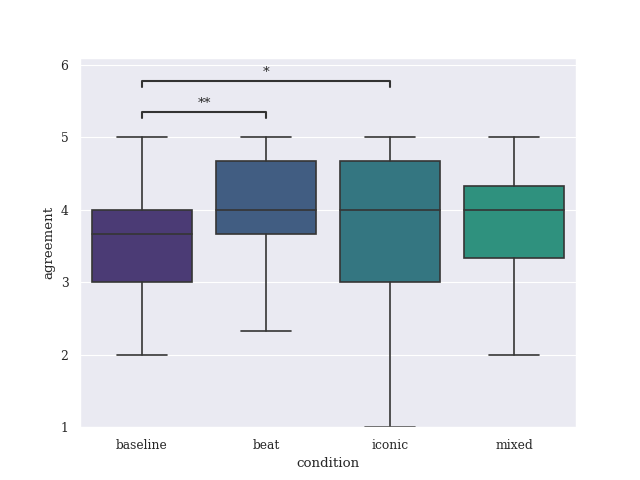}
  \caption{Engagement}\label{fig:engagement}
\endminipage
\minipage{0.25\textwidth}%
  \includegraphics[width=\linewidth]{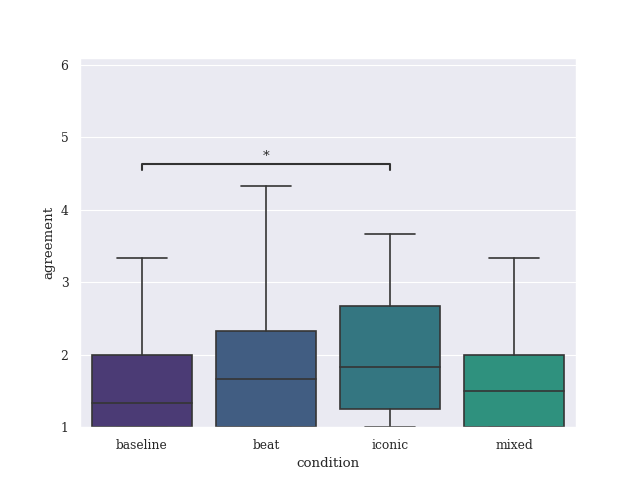}
  \caption{Acceptance}\label{fig:acceptance}
\endminipage
\end{figure}
Next to the objective understanding, the users had to answer a questionnaire on their subjective perception of the interaction. This interaction questionnaire is a selection of 14 dimensions taken from the Artificial Social Agents Questionnaire (ASAQ) questionnaire \cite{fitrianie_artificial-social-agent_2022}, and two additional, more explanation-related dimensions, the subjective understanding and the connection between agent and understanding \footnote{for the full questionnaire see Appendix \ref{sec:appendixSQ}}. It addresses social perception and communicative effects. Each of these 16 dimensions was tested with three different statements and were answered by the participant using a 5-point Likert scale.

\begin{figure}[!t]
    \centering
    \includegraphics[width=\columnwidth]{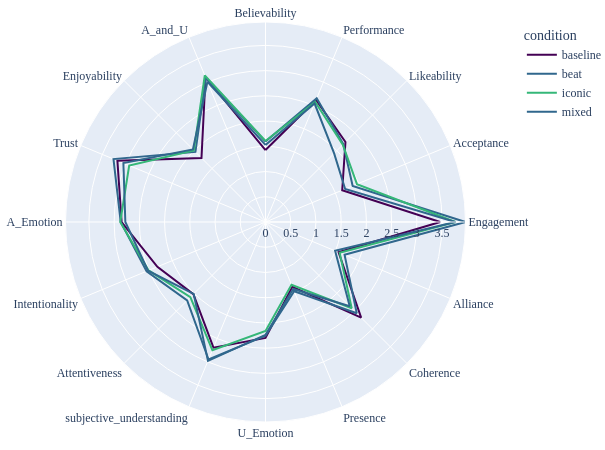}
    \caption{Radar chart comparing the four gesture conditions}
    \label{fig:radar}
\end{figure}

When comparing how the different interactions were perceived by the user, the conditions do not show substantial differences (Fig. \ref{fig:radar}). We only see a significant difference in 4 out of the 16 dimensions. In the level of trust the users put into the agent, the beat condition significantly outperforms the iconic condition (U = 1.256e+03 p = 4.811e-02) (Fig.\ref{fig:Trust}). The user perceives the baseline condition to be more likable than the mixed condition (U = 1.404e+03 p = 3.841e-02) (Fig. \ref{fig:likeability}). When it comes to engagement, the beat (U = 7.135e+02 p = 4.379e-03) and the iconic condition (U = 7.650e+02 p = 3.143e-02) are perceived as more engaging than the baseline (Fig. \ref{fig:engagement}). Also, the acceptance and willingness to use the agent again are higher in the iconic than in the baseline condition (U = 7.810e+02 p =  3.982e-02) (Fig.\ref{fig:acceptance}).

\subsection{Discussion}

The results of the present study suggest that, in the context of a board game explanation delivered by an embodied virtual agent, the use of co-speech gestures does not necessarily improve understanding. On the contrary, the non-iconic conditions (baseline and beat) tended to result in better deep enabledness of the participants than the iconic conditions (iconic and mixed). At the same time, we do see an improvement in understanding in comparison to the SNAPE model. The fact that the embodied baseline condition exceeds the previous one (baseline23) by a significant margin can either be explained by differences in explanation quality or by the added embodiment of the agent. The former is possible as SNAPE can only produce one piece of information per utterance, while our model improves explanation quality by generating utterances with a more complex structure and a higher information density, whenever reasonable (Sect. \ref{sec:themodel}). We thus take these results as evidence of a positive impact of our model, even if the explanation was not adaptive. The latter is supported by the so-called "embodiment effect" \cite{mayer_embodiment_2012} showing better learning of materials presented with an embodied character. In any case, since the difference in understanding is between the current baseline condition and the baseline23 condition, it cannot be explained by the addition of gestures. 

A comparison of perceived interaction quality between the gesture conditions does not reveal many differences. Only four of the sixteen dimensions show a significant difference. While the performed gestures improve user engagement, they decrease the likability of the agent. When looking at the user's trust, the overall type of gesture seems to have a smaller impact than the fluidity of the performed gesture. An agent using beat gestures is perceived as more trustworthy while performing iconic gestures significantly increases the probability that a user will reuse the agent compared to the baseline condition.

In summary, hypotheses H1 and H2 must be rejected, while hypothesis H3 can be partially accepted. We see an improvement in perception in some dimensions when either beat or iconic gestures are used, but the combination never outperforms the baseline. We did not find a significantly better understanding of the mixed condition compared to any of the other conditions. In fact, the iconic condition produces a significantly worse understanding than the baseline. The extended agent outperforms the previous version, especially when looking at deep comprehension. However, it should be noted that the mixed and iconic conditions are the only gesture conditions that do not outperform the adaptive condition in general comprehension, and the difference between the adaptive condition and the baseline, beat, and iconic conditions is even greater than the difference between the mixed and adaptive conditions in deep comprehension.

\section{Conclusions}
\label{sec:discussion}
In this paper, we investigated how gestures affect users' perception and understanding of explanations given by an embodied agent. Using a modified version of the SNAPE explanation generation model, coupled with a graph-based gesture generation algorithm, we could systematically adjust gesture parameters and measure participants' comprehension. Our results show only slight differences in user perception and general understanding if confronted with different gesture types. Notably, our results reveal a significant decrease in deep understanding when presented with iconic gestures. Compared to previous research, these results are rather unexpected. Although previous research has shown that gestures can distract people and reduce performance \cite{kramer_effects_2003}, most studies suggested that gestures improve the quality of an interaction \cite{kopp_computational_2017, wu_effects_2014, van_merrienboer_cognitive_2005, dargue_when_2019, ping_hands_2008}. Looking at these findings, how can the observed effects be explained?

One important consideration is the amount of cognitive load that the stimuli place on the participants \cite{castro-alonso_learning_2014, marcus_should_2013, sweller_cognitive_1998}. In our study, participants get a verbal explanation of a complex board game in a short period of time, which is likely to incur a high cognitive load \cite{van_merrienboer_cognitive_2005} and thus limit participants' learning. Additionally, the automatically generated explanations may introduce artifacts that increase cognitive load even further  \cite{paris_linguistic_2000,johnsen_experiences_2005}. In this situation, the presentation of iconic gestures that need to be integrated with verbally conveyed content might have additionally taxed cognitive resources. This is in line with \textit{cognitive resource theory} which assumes a competition between modalities that need to be processed in parallel \cite{wickens_compatibility_1983} or when input from one modality needs to be transferred to a task that needs another modality. This is supported by the fact that the negative effects of iconic gestures are found in participant's deep understanding, but not in the general recall questions. We may thus conclude that the combination of semantically meaningful speech and gesture might have overwhelmed participants (as reported by \citet{dargue_not_2018}) and that the induced cognitive load \cite{woods_can_2002, fan_reducing_2006} might have prevented participants from internalizing the new information in the given time.  

In contrast, shallow understanding and deep enabledness are not hampered when explanations are given along with beat gestures. This observation is in line with the finding that beat gestures are more beneficial in difficult tasks than in simple ones \cite{gluhareva_training_2016}, as well as with the view that iconic and metaphoric gestures are not always more beneficial to comprehension than deictic or beat gestures \cite{dargue_when_2019}.

However, it is not possible to finally answer this question as the present study only examined redundant iconic gestures that re-instantiate information already conveyed verbally, albeit apparently sometimes in ways that may have led to confusion. It is generally acknowledged that gestures can enhance communication by providing additional information \cite{hostetter_when_2011}. This aligns with the concept of spreading information across multiple modalities to reduce cognitive load, as discussed above (Sect. \ref{sec:cognitiveload}). Thus, to gain further insight into the effects of an embodied agent's synthetic gestures on listeners' cognitive load, additional studies on redundant, supplementary, and complementary gestures are necessary. Further, there are multiple metrics available for measuring cognitive load \cite{sweller_measuring_2011}, which should be included in future studies to better delineate these effects.

Another aspect that goes beyond the scope of this paper is the extent to which the presented iconic gestures are familiar to the addressees. \citet{dargue_not_2018} found that a well-known gesture is easier to interpret and thus puts less load on the user. Whether this effect is found in the present board game explanations and the pre-recorded gestures will be tested in a follow-up study, measuring the familiarity with the used gestures. As for other limitations of our study, we only tested the effects of specific gestures in a specific type of interaction (an explanation) and a single domain (the board game Quarto!). This study hence cannot provide generalizable results and we are aware of studies showing different or even contrary findings. This again stresses the importance of shifting the perspective to representational gestures and learning more about the effects these gestures have or do not have in different kinds of interactions between human users and virtual multimodal agents. In addition to this, the study is only considering the effects of supplementary gestures on understanding and perceived interaction quality. By now, we cannot derive conclusions about the effects of representational gestures in explanations, which will be a research question for a follow-up study.

In summary, this study suggests that co-speech gestures do not necessarily improve explanations given by an embodied agent. When incorporating gestures, it is thus important to consider when to use which particular type of gesture, with which particular addressee, and at which particular point in the explanation process. The more meaningful and complex a gesture may seem, the more important this is. 

\bibliographystyle{ACM-Reference-Format}
\bibliography{references}


\begin{thebibliography}{102}


\ifx \showCODEN    \undefined \def \showCODEN     #1{\unskip}     \fi
\ifx \showDOI      \undefined \def \showDOI       #1{#1}\fi
\ifx \showISBNx    \undefined \def \showISBNx     #1{\unskip}     \fi
\ifx \showISBNxiii \undefined \def \showISBNxiii  #1{\unskip}     \fi
\ifx \showISSN     \undefined \def \showISSN      #1{\unskip}     \fi
\ifx \showLCCN     \undefined \def \showLCCN      #1{\unskip}     \fi
\ifx \shownote     \undefined \def \shownote      #1{#1}          \fi
\ifx \showarticletitle \undefined \def \showarticletitle #1{#1}   \fi
\ifx \showURL      \undefined \def \showURL       {\relax}        \fi
\providecommand\bibfield[2]{#2}
\providecommand\bibinfo[2]{#2}
\providecommand\natexlab[1]{#1}
\providecommand\showeprint[2][]{arXiv:#2}

\bibitem[Aussems and Kita(2019)]%
        {aussems_seeing_2019}
\bibfield{author}{\bibinfo{person}{Suzanne Aussems} {and} \bibinfo{person}{Sotaro Kita}.} \bibinfo{year}{2019}\natexlab{}.
\newblock \showarticletitle{Seeing {Iconic} {Gestures} {While} {Encoding} {Events} {Facilitates} {Children}'s {Memory} of {These} {Events}}.
\newblock \bibinfo{journal}{\emph{Child Development}} \bibinfo{volume}{90}, \bibinfo{number}{4} (\bibinfo{date}{July} \bibinfo{year}{2019}), \bibinfo{pages}{1123--1137}.
\newblock
\showISSN{0009-3920, 1467-8624}
\urldef\tempurl%
\url{https://doi.org/10.1111/cdev.12988}
\showDOI{\tempurl}


\bibitem[Baevski et~al\mbox{.}(2020)]%
        {baevski_wav2vec_2020}
\bibfield{author}{\bibinfo{person}{Alexei Baevski}, \bibinfo{person}{Henry Zhou}, \bibinfo{person}{Abdelrahman Mohamed}, {and} \bibinfo{person}{Michael Auli}.} \bibinfo{year}{2020}\natexlab{}.
\newblock \bibinfo{title}{wav2vec 2.0: {A} {Framework} for {Self}-{Supervised} {Learning} of {Speech} {Representations}}.
\newblock
\newblock
\urldef\tempurl%
\url{https://doi.org/10.48550/arXiv.2006.11477}
\showDOI{\tempurl}


\bibitem[Bailenson and Yee(2005)]%
        {bailenson_digital_2005}
\bibfield{author}{\bibinfo{person}{Jeremy~N. Bailenson} {and} \bibinfo{person}{Nick Yee}.} \bibinfo{year}{2005}\natexlab{}.
\newblock \showarticletitle{Digital {Chameleons}: {Automatic} {Assimilation} of {Nonverbal} {Gestures} in {Immersive} {Virtual} {Environments}}.
\newblock \bibinfo{journal}{\emph{Psychological Science}} \bibinfo{volume}{16}, \bibinfo{number}{10} (\bibinfo{date}{Oct.} \bibinfo{year}{2005}), \bibinfo{pages}{814--819}.
\newblock
\showISSN{0956-7976}
\urldef\tempurl%
\url{https://doi.org/10.1111/j.1467-9280.2005.01619.x}
\showDOI{\tempurl}
\newblock
\shownote{Publisher: SAGE Publications Inc}.


\bibitem[Belpaeme et~al\mbox{.}(2018)]%
        {belpaeme_guidelines_2018}
\bibfield{author}{\bibinfo{person}{Tony Belpaeme}, \bibinfo{person}{Paul Vogt}, \bibinfo{person}{Rianne van~den Berghe}, \bibinfo{person}{Kirsten Bergmann}, \bibinfo{person}{Tilbe Göksun}, \bibinfo{person}{Mirjam de Haas}, \bibinfo{person}{Junko Kanero}, \bibinfo{person}{James Kennedy}, \bibinfo{person}{Aylin~C. Küntay}, \bibinfo{person}{Ora Oudgenoeg-Paz}, \bibinfo{person}{Fotios Papadopoulos}, \bibinfo{person}{Thorsten Schodde}, \bibinfo{person}{Josje Verhagen}, \bibinfo{person}{Christopher~D. Wallbridge}, \bibinfo{person}{Bram Willemsen}, \bibinfo{person}{Jan de Wit}, \bibinfo{person}{Vasfiye Geçkin}, \bibinfo{person}{Laura Hoffmann}, \bibinfo{person}{Stefan Kopp}, \bibinfo{person}{Emiel Krahmer}, \bibinfo{person}{Ezgi Mamus}, \bibinfo{person}{Jean-Marc Montanier}, \bibinfo{person}{Cansu Oranç}, {and} \bibinfo{person}{Amit~Kumar Pandey}.} \bibinfo{year}{2018}\natexlab{}.
\newblock \showarticletitle{Guidelines for {Designing} {Social} {Robots} as {Second} {Language} {Tutors}}.
\newblock \bibinfo{journal}{\emph{International Journal of Social Robotics}} \bibinfo{volume}{10}, \bibinfo{number}{3} (\bibinfo{date}{June} \bibinfo{year}{2018}), \bibinfo{pages}{325--341}.
\newblock
\showISSN{1875-4805}
\urldef\tempurl%
\url{https://doi.org/10.1007/s12369-018-0467-6}
\showDOI{\tempurl}


\bibitem[Bergmann and Kopp(2009)]%
        {bergmann_increasing_2009}
\bibfield{author}{\bibinfo{person}{Kirsten Bergmann} {and} \bibinfo{person}{Stefan Kopp}.} \bibinfo{year}{2009}\natexlab{}.
\newblock \showarticletitle{Increasing the {Expressiveness} of {Virtual} {Agents}– {Autonomous} {Generation} of {Speech} and {Gesture} for {Spatial} {Description} {Tasks}}. In \bibinfo{booktitle}{\emph{Proceedings of the 8th {International} {Joint} {Conference} on {Autonomous} {Agents} and {Multiagent} {Systems}}}.
\newblock


\bibitem[Bergmann et~al\mbox{.}(2010)]%
        {bergmann_individualized_2010}
\bibfield{author}{\bibinfo{person}{Kirsten Bergmann}, \bibinfo{person}{Stefan Kopp}, {and} \bibinfo{person}{Friederike Eyssel}.} \bibinfo{year}{2010}\natexlab{}.
\newblock \showarticletitle{Individualized {Gesturing} {Outperforms} {Average} {Gesturing} – {Evaluating} {Gesture} {Production} in {Virtual} {Humans}}. In \bibinfo{booktitle}{\emph{Proceedings of the 10th {International} {Conference} on {Intelligent} {Virtual} {Agents}}}, \bibfield{editor}{\bibinfo{person}{Jan Allbeck}, \bibinfo{person}{Norman Badler}, \bibinfo{person}{Timothy Bickmore}, \bibinfo{person}{Catherine Pelachaud}, {and} \bibinfo{person}{Alla Safonova}} (Eds.). \bibinfo{publisher}{Springer}, \bibinfo{address}{Berlin, Heidelberg}, \bibinfo{pages}{104--117}.
\newblock
\showISBNx{978-3-642-15892-6}
\urldef\tempurl%
\url{https://doi.org/10.1007/978-3-642-15892-6_11}
\showDOI{\tempurl}


\bibitem[Bergmann and Macedonia(2013)]%
        {bergmann_virtual_2013}
\bibfield{author}{\bibinfo{person}{Kirsten Bergmann} {and} \bibinfo{person}{Manuela Macedonia}.} \bibinfo{year}{2013}\natexlab{}.
\newblock \showarticletitle{A {Virtual} {Agent} as {Vocabulary} {Trainer}: {Iconic} {Gestures} {Help} to {Improve} {Learners}’ {Memory} {Performance}}. In \bibinfo{booktitle}{\emph{Proceedings of the 13th {International} {Conference} on {Intelligent} {Virtual} {Agents}}}, \bibfield{editor}{\bibinfo{person}{Ruth Aylett}, \bibinfo{person}{Brigitte Krenn}, \bibinfo{person}{Catherine Pelachaud}, {and} \bibinfo{person}{Hiroshi Shimodaira}} (Eds.). \bibinfo{publisher}{Springer}, \bibinfo{address}{Berlin, Heidelberg}, \bibinfo{pages}{139--148}.
\newblock
\showISBNx{978-3-642-40415-3}
\urldef\tempurl%
\url{https://doi.org/10.1007/978-3-642-40415-3_12}
\showDOI{\tempurl}


\bibitem[Bojanowski et~al\mbox{.}(2017)]%
        {bojanowski_enriching_2017}
\bibfield{author}{\bibinfo{person}{Piotr Bojanowski}, \bibinfo{person}{Edouard Grave}, \bibinfo{person}{Armand Joulin}, {and} \bibinfo{person}{Tomas Mikolov}.} \bibinfo{year}{2017}\natexlab{}.
\newblock \bibinfo{title}{Enriching {Word} {Vectors} with {Subword} {Information}}.
\newblock
\newblock
\urldef\tempurl%
\url{https://doi.org/10.48550/arXiv.1607.04606}
\showDOI{\tempurl}
\newblock
\shownote{arXiv:1607.04606 [cs]}.


\bibitem[Bosker and Peeters(2021)]%
        {bosker_beat_2021}
\bibfield{author}{\bibinfo{person}{Hans~Rutger Bosker} {and} \bibinfo{person}{David Peeters}.} \bibinfo{year}{2021}\natexlab{}.
\newblock \showarticletitle{Beat gestures influence which speech sounds you hear}.
\newblock \bibinfo{journal}{\emph{Proceedings of the Royal Society B: Biological Sciences}} \bibinfo{volume}{288}, \bibinfo{number}{1943} (\bibinfo{date}{Jan.} \bibinfo{year}{2021}), \bibinfo{pages}{20202419}.
\newblock
\urldef\tempurl%
\url{https://doi.org/10.1098/rspb.2020.2419}
\showDOI{\tempurl}
\newblock
\shownote{Publisher: Royal Society}.


\bibitem[Breckinridge~Church et~al\mbox{.}(2007)]%
        {breckinridge_church_role_2007}
\bibfield{author}{\bibinfo{person}{Ruth Breckinridge~Church}, \bibinfo{person}{Philip Garber}, {and} \bibinfo{person}{Kathryn Rogalski}.} \bibinfo{year}{2007}\natexlab{}.
\newblock \showarticletitle{The role of gesture in memory and social communication}.
\newblock \bibinfo{journal}{\emph{Gesture}} \bibinfo{volume}{7}, \bibinfo{number}{2} (\bibinfo{date}{June} \bibinfo{year}{2007}), \bibinfo{pages}{137--158}.
\newblock
\showISSN{1568-1475, 1569-9773}
\urldef\tempurl%
\url{https://doi.org/10.1075/gest.7.2.02bre}
\showDOI{\tempurl}


\bibitem[Buschmeier et~al\mbox{.}(2023)]%
        {buschmeier_forms_2023}
\bibfield{author}{\bibinfo{person}{Hendrik Buschmeier}, \bibinfo{person}{Heike~M. Buhl}, \bibinfo{person}{Friederike Kern}, \bibinfo{person}{Angela Grimminger}, \bibinfo{person}{Helen Beierling}, \bibinfo{person}{Josephine Fisher}, \bibinfo{person}{André Groß}, \bibinfo{person}{Ilona Horwath}, \bibinfo{person}{Nils Klowait}, \bibinfo{person}{Stefan Lazarov}, \bibinfo{person}{Michael Lenke}, \bibinfo{person}{Vivien Lohmer}, \bibinfo{person}{Katharina Rohlfing}, \bibinfo{person}{Ingrid Scharlau}, \bibinfo{person}{Amit Singh}, \bibinfo{person}{Lutz Terfloth}, \bibinfo{person}{Anna-Lisa Vollmer}, \bibinfo{person}{Yu Wang}, \bibinfo{person}{Annedore Wilmes}, {and} \bibinfo{person}{Britta Wrede}.} \bibinfo{year}{2023}\natexlab{}.
\newblock \bibinfo{title}{Forms of {Understanding} of {XAI}-{Explanations}}.
\newblock
\newblock
\urldef\tempurl%
\url{http://arxiv.org/abs/2311.08760}
\showURL{%
\tempurl}
\newblock
\shownote{arXiv:2311.08760 [cs]}.


\bibitem[Cassell et~al\mbox{.}(1999)]%
        {cassell_speech-gesture_1999}
\bibfield{author}{\bibinfo{person}{Justine Cassell}, \bibinfo{person}{David McNeill}, {and} \bibinfo{person}{Karl-Erik McCullough}.} \bibinfo{year}{1999}\natexlab{}.
\newblock \showarticletitle{Speech-gesture mismatches: {Evidence} for one underlying representation of linguistic and nonlinguistic information}.
\newblock \bibinfo{journal}{\emph{Pragmatics \& cognition}} \bibinfo{volume}{7}, \bibinfo{number}{1} (\bibinfo{year}{1999}), \bibinfo{pages}{1--34}.
\newblock
\urldef\tempurl%
\url{https://doi.org/10.1075/pc.7.1.03cas}
\showDOI{\tempurl}
\newblock
\shownote{Publisher: John Benjamins}.


\bibitem[Cassell et~al\mbox{.}(1994)]%
        {cassell_animated_1994}
\bibfield{author}{\bibinfo{person}{Justine Cassell}, \bibinfo{person}{Catherine Pelachaud}, \bibinfo{person}{Norman Badler}, \bibinfo{person}{Mark Steedman}, \bibinfo{person}{Brett Achorn}, \bibinfo{person}{Tripp Becket}, \bibinfo{person}{Brett Douville}, \bibinfo{person}{Scott Prevost}, {and} \bibinfo{person}{Matthew Stone}.} \bibinfo{year}{1994}\natexlab{}.
\newblock \showarticletitle{Animated conversation: rule-based generation of facial expression, gesture \& spoken intonation for multiple conversational agents}. In \bibinfo{booktitle}{\emph{Proceedings of the 21st annual conference on {Computer} graphics and interactive techniques}} \emph{(\bibinfo{series}{{SIGGRAPH} '94})}. \bibinfo{publisher}{Association for Computing Machinery}, \bibinfo{address}{New York, NY, USA}, \bibinfo{pages}{413--420}.
\newblock
\showISBNx{978-0-89791-667-7}
\urldef\tempurl%
\url{https://doi.org/10.1145/192161.192272}
\showDOI{\tempurl}


\bibitem[Cassell et~al\mbox{.}(2004)]%
        {cassell_beat_2004}
\bibfield{author}{\bibinfo{person}{Justine Cassell}, \bibinfo{person}{Hannes~Högni Vilhjálmsson}, {and} \bibinfo{person}{Timothy Bickmore}.} \bibinfo{year}{2004}\natexlab{}.
\newblock \showarticletitle{{BEAT}: the {Behavior} {Expression} {Animation} {Toolkit}}.
\newblock In \bibinfo{booktitle}{\emph{Life-{Like} {Characters}: {Tools}, {Affective} {Functions}, and {Applications}}}, \bibfield{editor}{\bibinfo{person}{Helmut Prendinger} {and} \bibinfo{person}{Mitsuru Ishizuka}} (Eds.). \bibinfo{publisher}{Springer}, \bibinfo{address}{Berlin, Heidelberg}, \bibinfo{pages}{163--185}.
\newblock
\showISBNx{978-3-662-08373-4}
\urldef\tempurl%
\url{https://doi.org/10.1007/978-3-662-08373-4_8}
\showDOI{\tempurl}


\bibitem[Castellano(2024)]%
        {castellano_home_2024}
\bibfield{author}{\bibinfo{person}{Brandon Castellano}.} \bibinfo{year}{2024}\natexlab{}.
\newblock \bibinfo{title}{Home - {PySceneDetect}}.
\newblock
\newblock
\urldef\tempurl%
\url{https://www.scenedetect.com/}
\showURL{%
\tempurl}


\bibitem[Castro-Alonso et~al\mbox{.}(2014)]%
        {castro-alonso_learning_2014}
\bibfield{author}{\bibinfo{person}{Juan~C. Castro-Alonso}, \bibinfo{person}{Paul Ayres}, {and} \bibinfo{person}{Fred Paas}.} \bibinfo{year}{2014}\natexlab{}.
\newblock \showarticletitle{Learning from observing hands in static and animated versions of non-manipulative tasks}.
\newblock \bibinfo{journal}{\emph{Learning and Instruction}}  \bibinfo{volume}{34} (\bibinfo{date}{Dec.} \bibinfo{year}{2014}), \bibinfo{pages}{11--21}.
\newblock
\showISSN{09594752}
\urldef\tempurl%
\url{https://doi.org/10.1016/j.learninstruc.2014.07.005}
\showDOI{\tempurl}


\bibitem[Chen et~al\mbox{.}(2012)]%
        {chen_multimodal_2012}
\bibfield{author}{\bibinfo{person}{Fang Chen}, \bibinfo{person}{Natalie Ruiz}, \bibinfo{person}{Eric Choi}, \bibinfo{person}{Julien Epps}, \bibinfo{person}{M.~Asif Khawaja}, \bibinfo{person}{Ronnie Taib}, \bibinfo{person}{Bo Yin}, {and} \bibinfo{person}{Yang Wang}.} \bibinfo{year}{2012}\natexlab{}.
\newblock \showarticletitle{Multimodal behavior and interaction as indicators of cognitive load}.
\newblock \bibinfo{journal}{\emph{ACM Transactions on Interactive Intelligent Systems}} \bibinfo{volume}{2}, \bibinfo{number}{4} (\bibinfo{date}{Dec.} \bibinfo{year}{2012}), \bibinfo{pages}{1--36}.
\newblock
\showISSN{2160-6455, 2160-6463}
\urldef\tempurl%
\url{https://doi.org/10.1145/2395123.2395127}
\showDOI{\tempurl}


\bibitem[Cherti et~al\mbox{.}(2022)]%
        {cherti_reproducible_2022}
\bibfield{author}{\bibinfo{person}{Mehdi Cherti}, \bibinfo{person}{Romain Beaumont}, \bibinfo{person}{Ross Wightman}, \bibinfo{person}{Mitchell Wortsman}, \bibinfo{person}{Gabriel Ilharco}, \bibinfo{person}{Cade Gordon}, \bibinfo{person}{Christoph Schuhmann}, \bibinfo{person}{Ludwig Schmidt}, {and} \bibinfo{person}{Jenia Jitsev}.} \bibinfo{year}{2022}\natexlab{}.
\newblock \bibinfo{title}{Reproducible scaling laws for contrastive language-image learning}.
\newblock
\newblock
\urldef\tempurl%
\url{http://arxiv.org/abs/2212.07143}
\showURL{%
\tempurl}
\newblock
\shownote{arXiv:2212.07143 [cs]}.


\bibitem[Community(2018)]%
        {community_blender_2018}
\bibfield{author}{\bibinfo{person}{Blender~Online Community}.} \bibinfo{year}{2018}\natexlab{}.
\newblock \bibinfo{booktitle}{\emph{Blender - a {3D} modelling and rendering package}}.
\newblock \bibinfo{publisher}{Blender Foundation}, \bibinfo{address}{Stichting Blender Foundation, Amsterdam}.
\newblock
\urldef\tempurl%
\url{http://www.blender.org}
\showURL{%
\tempurl}


\bibitem[Dargue and Sweller(2018)]%
        {dargue_not_2018}
\bibfield{author}{\bibinfo{person}{Nicole Dargue} {and} \bibinfo{person}{Naomi Sweller}.} \bibinfo{year}{2018}\natexlab{}.
\newblock \showarticletitle{Not {All} {Gestures} are {Created} {Equal}: {The} {Effects} of {Typical} and {Atypical} {Iconic} {Gestures} on {Narrative} {Comprehension}}.
\newblock \bibinfo{journal}{\emph{Journal of Nonverbal Behavior}}  \bibinfo{volume}{42} (\bibinfo{date}{Sept.} \bibinfo{year}{2018}), \bibinfo{pages}{1--19}.
\newblock
\urldef\tempurl%
\url{https://doi.org/10.1007/s10919-018-0278-3}
\showDOI{\tempurl}


\bibitem[Dargue et~al\mbox{.}(2019)]%
        {dargue_when_2019}
\bibfield{author}{\bibinfo{person}{Nicole Dargue}, \bibinfo{person}{Naomi Sweller}, {and} \bibinfo{person}{Michael Jones}.} \bibinfo{year}{2019}\natexlab{}.
\newblock \showarticletitle{When {Our} {Hands} {Help} {Us} {Understand}: {A} {Meta}-{Analysis} {Into} the {Effects} of {Gesture} on {Comprehension}}.
\newblock \bibinfo{journal}{\emph{Psychological Bulletin}}  \bibinfo{volume}{145} (\bibinfo{date}{June} \bibinfo{year}{2019}).
\newblock
\urldef\tempurl%
\url{https://doi.org/10.1037/bul0000202}
\showDOI{\tempurl}


\bibitem[Davis(2018)]%
        {davis_impact_2018}
\bibfield{author}{\bibinfo{person}{Robert~O. Davis}.} \bibinfo{year}{2018}\natexlab{}.
\newblock \showarticletitle{The impact of pedagogical agent gesturing in multimedia learning environments: {A} meta-analysis}.
\newblock \bibinfo{journal}{\emph{Educational Research Review}} (\bibinfo{year}{2018}).
\newblock
\urldef\tempurl%
\url{https://doi.org/10.1016/J.EDUREV.2018.05.002}
\showDOI{\tempurl}


\bibitem[de~Wit et~al\mbox{.}(2018)]%
        {de_wit_effect_2018}
\bibfield{author}{\bibinfo{person}{Jan de Wit}, \bibinfo{person}{Thorsten Schodde}, \bibinfo{person}{Bram Willemsen}, \bibinfo{person}{Kirsten Bergmann}, \bibinfo{person}{Mirjam de Haas}, \bibinfo{person}{Stefan Kopp}, \bibinfo{person}{Emiel Krahmer}, {and} \bibinfo{person}{Paul Vogt}.} \bibinfo{year}{2018}\natexlab{}.
\newblock \showarticletitle{The {Effect} of a {Robot}'s {Gestures} and {Adaptive} {Tutoring} on {Children}'s {Acquisition} of {Second} {Language} {Vocabularies}}. In \bibinfo{booktitle}{\emph{Proceedings of the 2018 {ACM}/{IEEE} {International} {Conference} on {Human}-{Robot} {Interaction}}} \emph{(\bibinfo{series}{{HRI} '18})}. \bibinfo{publisher}{Association for Computing Machinery}, \bibinfo{address}{New York, NY, USA}, \bibinfo{pages}{50--58}.
\newblock
\showISBNx{978-1-4503-4953-6}
\urldef\tempurl%
\url{https://doi.org/10.1145/3171221.3171277}
\showDOI{\tempurl}


\bibitem[Fan and Lei(2006)]%
        {fan_reducing_2006}
\bibfield{author}{\bibinfo{person}{Lisa Fan} {and} \bibinfo{person}{Minxiao Lei}.} \bibinfo{year}{2006}\natexlab{}.
\newblock \showarticletitle{Reducing {Cognitive} {Overload} by {Meta}-{Learning} {Assisted} {Algorithm} {Selection}}. In \bibinfo{booktitle}{\emph{2006 5th {IEEE} {International} {Conference} on {Cognitive} {Informatics}}}, Vol.~\bibinfo{volume}{1}. \bibinfo{pages}{120--125}.
\newblock
\urldef\tempurl%
\url{https://doi.org/10.1109/COGINF.2006.365686}
\showDOI{\tempurl}


\bibitem[Fisher et~al\mbox{.}(2023)]%
        {fisher_exploring_2023}
\bibfield{author}{\bibinfo{person}{Josephine~B Fisher}, \bibinfo{person}{Amelie~S Robrecht}, \bibinfo{person}{Stefan Kopp}, {and} \bibinfo{person}{Katharina~J Rohlfing}.} \bibinfo{year}{2023}\natexlab{}.
\newblock \showarticletitle{Exploring the {Semantic} {Dialogue} {Patterns} of {Explanations} – a {Case} {Study} of {Game} {Explanations}}. In \bibinfo{booktitle}{\emph{Proceedings of the 27th {Workshop} on the {Semantics} and {Pragmatics} of {Dialogue} ({SemDial} 2023)}}.
\newblock


\bibitem[Fitrianie et~al\mbox{.}(2022)]%
        {fitrianie_artificial-social-agent_2022}
\bibfield{author}{\bibinfo{person}{Siska Fitrianie}, \bibinfo{person}{Merijn Bruijnes}, \bibinfo{person}{Fengxiang Li}, \bibinfo{person}{Amal Abdulrahman}, {and} \bibinfo{person}{Willem-Paul Brinkman}.} \bibinfo{year}{2022}\natexlab{}.
\newblock \showarticletitle{The artificial-social-agent questionnaire: establishing the long and short questionnaire versions}. In \bibinfo{booktitle}{\emph{Proceedings of the 22nd {ACM} {International} {Conference} on {Intelligent} {Virtual} {Agents}}}. \bibinfo{publisher}{ACM}, \bibinfo{address}{Faro Portugal}, \bibinfo{pages}{1--8}.
\newblock
\showISBNx{978-1-4503-9248-8}
\urldef\tempurl%
\url{https://doi.org/10.1145/3514197.3549612}
\showDOI{\tempurl}


\bibitem[foundation(2024a)]%
        {ted_foundation_ted_2024}
\bibfield{author}{\bibinfo{person}{TED foundation}.} \bibinfo{year}{2024}\natexlab{a}.
\newblock \bibinfo{title}{{TED} - {YouTube}}.
\newblock
\newblock
\urldef\tempurl%
\url{https://www.youtube.com/channel/UCAuUUnT6oDeKwE6v1NGQxug}
\showURL{%
\tempurl}


\bibitem[foundation(2024b)]%
        {ted_foundation_tedx_2024}
\bibfield{author}{\bibinfo{person}{TED foundation}.} \bibinfo{year}{2024}\natexlab{b}.
\newblock \bibinfo{title}{{TEDx} {Talks} - {YouTube}}.
\newblock
\newblock
\urldef\tempurl%
\url{https://www.youtube.com/user/tedxtalks}
\showURL{%
\tempurl}


\bibitem[Gadre et~al\mbox{.}(2023)]%
        {gadre_datacomp_2023}
\bibfield{author}{\bibinfo{person}{Samir~Yitzhak Gadre}, \bibinfo{person}{Gabriel Ilharco}, \bibinfo{person}{Alex Fang}, \bibinfo{person}{Jonathan Hayase}, \bibinfo{person}{Georgios Smyrnis}, \bibinfo{person}{Thao Nguyen}, \bibinfo{person}{Ryan Marten}, \bibinfo{person}{Mitchell Wortsman}, \bibinfo{person}{Dhruba Ghosh}, \bibinfo{person}{Jieyu Zhang}, \bibinfo{person}{Eyal Orgad}, \bibinfo{person}{Rahim Entezari}, \bibinfo{person}{Giannis Daras}, \bibinfo{person}{Sarah Pratt}, \bibinfo{person}{Vivek Ramanujan}, \bibinfo{person}{Yonatan Bitton}, \bibinfo{person}{Kalyani Marathe}, \bibinfo{person}{Stephen Mussmann}, \bibinfo{person}{Richard Vencu}, \bibinfo{person}{Mehdi Cherti}, \bibinfo{person}{Ranjay Krishna}, \bibinfo{person}{Pang~Wei Koh}, \bibinfo{person}{Olga Saukh}, \bibinfo{person}{Alexander Ratner}, \bibinfo{person}{Shuran Song}, \bibinfo{person}{Hannaneh Hajishirzi}, \bibinfo{person}{Ali Farhadi}, \bibinfo{person}{Romain Beaumont}, \bibinfo{person}{Sewoong Oh}, \bibinfo{person}{Alex Dimakis},
  \bibinfo{person}{Jenia Jitsev}, \bibinfo{person}{Yair Carmon}, \bibinfo{person}{Vaishaal Shankar}, {and} \bibinfo{person}{Ludwig Schmidt}.} \bibinfo{year}{2023}\natexlab{}.
\newblock \bibinfo{title}{{DataComp}: {In} search of the next generation of multimodal datasets}.
\newblock
\newblock
\urldef\tempurl%
\url{https://doi.org/10.48550/arXiv.2304.14108}
\showDOI{\tempurl}
\newblock
\shownote{arXiv:2304.14108 [cs]}.


\bibitem[Gluhareva and Prieto(2016)]%
        {gluhareva_training_2016}
\bibfield{author}{\bibinfo{person}{Daria Gluhareva} {and} \bibinfo{person}{Pilar Prieto}.} \bibinfo{year}{2016}\natexlab{}.
\newblock \showarticletitle{Training with rhythmic beat gestures benefits {L2} pronunciation in discourse-demanding situations}.
\newblock \bibinfo{journal}{\emph{Language Teaching Research}}  \bibinfo{volume}{21} (\bibinfo{date}{June} \bibinfo{year}{2016}).
\newblock
\urldef\tempurl%
\url{https://doi.org/10.1177/1362168816651463}
\showDOI{\tempurl}


\bibitem[Goldin-Meadow et~al\mbox{.}(2009)]%
        {goldin-meadow_gesturing_2009}
\bibfield{author}{\bibinfo{person}{Susan Goldin-Meadow}, \bibinfo{person}{Susan~Wagner Cook}, {and} \bibinfo{person}{Zachary~A. Mitchell}.} \bibinfo{year}{2009}\natexlab{}.
\newblock \showarticletitle{Gesturing {Gives} {Children} {New} {Ideas} {About} {Math}}.
\newblock \bibinfo{journal}{\emph{Psychological Science}} \bibinfo{volume}{20}, \bibinfo{number}{3} (\bibinfo{date}{March} \bibinfo{year}{2009}), \bibinfo{pages}{267--272}.
\newblock
\showISSN{0956-7976, 1467-9280}
\urldef\tempurl%
\url{https://doi.org/10.1111/j.1467-9280.2009.02297.x}
\showDOI{\tempurl}


\bibitem[Gratch et~al\mbox{.}(2007)]%
        {gratch_creating_2007}
\bibfield{author}{\bibinfo{person}{Jonathan Gratch}, \bibinfo{person}{Ning Wang}, \bibinfo{person}{Jillian Gerten}, \bibinfo{person}{Edward Fast}, {and} \bibinfo{person}{Robin Duffy}.} \bibinfo{year}{2007}\natexlab{}.
\newblock \showarticletitle{Creating {Rapport} with {Virtual} {Agents}}. In \bibinfo{booktitle}{\emph{Prodeedings of the 7th {International} {Conference} in {Intelligent} {Virtual} {Agents}}}, \bibfield{editor}{\bibinfo{person}{Catherine Pelachaud}, \bibinfo{person}{Jean-Claude Martin}, \bibinfo{person}{Elisabeth André}, \bibinfo{person}{Gérard Chollet}, \bibinfo{person}{Kostas Karpouzis}, {and} \bibinfo{person}{Danielle Pelé}} (Eds.). \bibinfo{publisher}{Springer}, \bibinfo{address}{Berlin, Heidelberg}, \bibinfo{pages}{125--138}.
\newblock
\showISBNx{978-3-540-74997-4}
\urldef\tempurl%
\url{https://doi.org/10.1007/978-3-540-74997-4_12}
\showDOI{\tempurl}


\bibitem[Group(2024)]%
        {group_postgresql_2024}
\bibfield{author}{\bibinfo{person}{PostgreSQL Global~Development Group}.} \bibinfo{year}{2024}\natexlab{}.
\newblock \bibinfo{title}{{PostgreSQL}}.
\newblock
\newblock
\urldef\tempurl%
\url{https://www.postgresql.org/}
\showURL{%
\tempurl}


\bibitem[Hostetter(2011)]%
        {hostetter_when_2011}
\bibfield{author}{\bibinfo{person}{Autumn~B. Hostetter}.} \bibinfo{year}{2011}\natexlab{}.
\newblock \showarticletitle{When do gestures communicate? {A} meta-analysis}.
\newblock \bibinfo{journal}{\emph{Psychological Bulletin}} \bibinfo{volume}{137}, \bibinfo{number}{2} (\bibinfo{date}{March} \bibinfo{year}{2011}), \bibinfo{pages}{297--315}.
\newblock
\showISSN{1939-1455}
\urldef\tempurl%
\url{https://doi.org/10.1037/a0022128}
\showDOI{\tempurl}


\bibitem[Hostetter and Bahl(2023)]%
        {hostetter_comparing_2023}
\bibfield{author}{\bibinfo{person}{Autumn~B. Hostetter} {and} \bibinfo{person}{Sonal Bahl}.} \bibinfo{year}{2023}\natexlab{}.
\newblock \showarticletitle{Comparing the cognitive load of gesture and action production: a dual-task study}.
\newblock \bibinfo{journal}{\emph{Language and Cognition}} \bibinfo{volume}{15}, \bibinfo{number}{3} (\bibinfo{date}{Sept.} \bibinfo{year}{2023}), \bibinfo{pages}{601--621}.
\newblock
\showISSN{1866-9808, 1866-9859}
\urldef\tempurl%
\url{https://doi.org/10.1017/langcog.2023.23}
\showDOI{\tempurl}


\bibitem[Johnsen et~al\mbox{.}(2005)]%
        {johnsen_experiences_2005}
\bibfield{author}{\bibinfo{person}{K. Johnsen}, \bibinfo{person}{R. Dickerson}, \bibinfo{person}{A. Raij}, \bibinfo{person}{B. Lok}, \bibinfo{person}{J. Jackson}, \bibinfo{person}{Min Shin}, \bibinfo{person}{J. Hernandez}, \bibinfo{person}{A. Stevens}, {and} \bibinfo{person}{D.S. Lind}.} \bibinfo{year}{2005}\natexlab{}.
\newblock \showarticletitle{Experiences in using immersive virtual characters to educate medical communication skills}. In \bibinfo{booktitle}{\emph{{IEEE} {Proceedings}. {VR} 2005. {Virtual} {Reality}, 2005.}} \bibinfo{pages}{179--186}.
\newblock
\urldef\tempurl%
\url{https://doi.org/10.1109/VR.2005.1492772}
\showDOI{\tempurl}
\newblock
\shownote{ISSN: 2375-5334}.


\bibitem[Johnstone(1994)]%
        {johnstone_repetition_1994}
\bibfield{editor}{\bibinfo{person}{Barbara Johnstone}} (Ed.). \bibinfo{year}{1994}\natexlab{}.
\newblock \bibinfo{booktitle}{\emph{Repetition in discourse: interdisciplinary perspectives}}.
\newblock Number v.47-48 in \bibinfo{series}{Advances in discourse processes}. \bibinfo{publisher}{Ablex Pub. Co}, \bibinfo{address}{Norwood, N.J}.
\newblock
\showISBNx{978-0-89391-830-9 978-0-89391-931-3 978-0-89391-831-6 978-0-89391-932-0}


\bibitem[Joulin et~al\mbox{.}(2016)]%
        {joulin_bag_2016}
\bibfield{author}{\bibinfo{person}{Armand Joulin}, \bibinfo{person}{Edouard Grave}, \bibinfo{person}{Piotr Bojanowski}, {and} \bibinfo{person}{Tomas Mikolov}.} \bibinfo{year}{2016}\natexlab{}.
\newblock \bibinfo{title}{Bag of {Tricks} for {Efficient} {Text} {Classification}}.
\newblock
\newblock
\urldef\tempurl%
\url{https://doi.org/10.48550/arXiv.1607.01759}
\showDOI{\tempurl}
\newblock
\shownote{arXiv:1607.01759 [cs]}.


\bibitem[Kane(2024)]%
        {kane_pgvectorpgvector_2024}
\bibfield{author}{\bibinfo{person}{Andrew Kane}.} \bibinfo{year}{2024}\natexlab{}.
\newblock \bibinfo{title}{pgvector/pgvector}.
\newblock
\newblock
\urldef\tempurl%
\url{https://github.com/pgvector/pgvector}
\showURL{%
\tempurl}
\newblock
\shownote{original-date: 2021-04-20T21:13:52Z}.


\bibitem[Kelly et~al\mbox{.}(2004)]%
        {kelly_neural_2004}
\bibfield{author}{\bibinfo{person}{S. Kelly}, \bibinfo{person}{Corinne Kravitz}, {and} \bibinfo{person}{Michael Hopkins}.} \bibinfo{year}{2004}\natexlab{}.
\newblock \showarticletitle{Neural correlates of bimodal speech and gesture comprehension}.
\newblock \bibinfo{journal}{\emph{Brain and Language}}  \bibinfo{volume}{89} (\bibinfo{year}{2004}), \bibinfo{pages}{253--260}.
\newblock
\urldef\tempurl%
\url{https://doi.org/10.1016/S0093-934X(03)00335-3}
\showDOI{\tempurl}


\bibitem[Kingma and Welling(2022)]%
        {kingma_auto-encoding_2022}
\bibfield{author}{\bibinfo{person}{Diederik~P. Kingma} {and} \bibinfo{person}{Max Welling}.} \bibinfo{year}{2022}\natexlab{}.
\newblock \bibinfo{title}{Auto-{Encoding} {Variational} {Bayes}}.
\newblock
\newblock
\urldef\tempurl%
\url{https://doi.org/10.48550/arXiv.1312.6114}
\showDOI{\tempurl}
\newblock
\shownote{arXiv:1312.6114 [cs, stat]}.


\bibitem[Kita and Özyürek(2003)]%
        {kita_what_2003}
\bibfield{author}{\bibinfo{person}{Sotaro Kita} {and} \bibinfo{person}{Asli Özyürek}.} \bibinfo{year}{2003}\natexlab{}.
\newblock \showarticletitle{What does cross-linguistic variation in semantic coordination of speech and gesture reveal?: {Evidence} for an interface representation of spatial thinking and speaking}.
\newblock \bibinfo{journal}{\emph{Journal of Memory and Language}} \bibinfo{volume}{48}, \bibinfo{number}{1} (\bibinfo{date}{Jan.} \bibinfo{year}{2003}), \bibinfo{pages}{16--32}.
\newblock
\showISSN{0749596X}
\urldef\tempurl%
\url{https://doi.org/10.1016/S0749-596X(02)00505-3}
\showDOI{\tempurl}


\bibitem[Kopp(2017a)]%
        {church_chapter_2017}
\bibfield{author}{\bibinfo{person}{Stefan Kopp}.} \bibinfo{year}{2017}\natexlab{a}.
\newblock \showarticletitle{Chapter 12. {Computational} gesture research: {Studying} the functions of gesture in human-agent interaction}.
\newblock In \bibinfo{booktitle}{\emph{Gesture {Studies}}}, \bibfield{editor}{\bibinfo{person}{R.~Breckinridge Church}, \bibinfo{person}{Martha~W. Alibali}, {and} \bibinfo{person}{Spencer~D. Kelly}} (Eds.). Vol.~\bibinfo{volume}{7}. \bibinfo{publisher}{John Benjamins Publishing Company}, \bibinfo{address}{Amsterdam}, \bibinfo{pages}{267--284}.
\newblock
\showISBNx{978-90-272-2849-9 978-90-272-6577-7}
\urldef\tempurl%
\url{https://doi.org/10.1075/gs.7.13kop}
\showDOI{\tempurl}


\bibitem[Kopp(2017b)]%
        {kopp_computational_2017}
\bibfield{author}{\bibinfo{person}{Stefan Kopp}.} \bibinfo{year}{2017}\natexlab{b}.
\newblock \showarticletitle{Computational gesture research: {Studying} the functions of gesture in human-agent interaction}.
\newblock In \bibinfo{booktitle}{\emph{Why {Gesture}?}} \bibinfo{publisher}{John Benjamins}, \bibinfo{pages}{267--284}.
\newblock
\urldef\tempurl%
\url{https://www.jbe-platform.com/content/books/9789027265777-gs.7.13kop}
\showURL{%
\tempurl}


\bibitem[Kopp et~al\mbox{.}(2006)]%
        {kopp_towards_2006}
\bibfield{author}{\bibinfo{person}{Stefan Kopp}, \bibinfo{person}{Brigitte Krenn}, \bibinfo{person}{Stacy Marsella}, \bibinfo{person}{Andrew Marshall}, \bibinfo{person}{Catherine Pelachaud}, \bibinfo{person}{Hannes Pirker}, \bibinfo{person}{Kristinn Thórisson}, {and} \bibinfo{person}{Hannes Vilhjálmsson}.} \bibinfo{year}{2006}\natexlab{}.
\newblock \showarticletitle{Towards a {Common} {Framework} for {Multimodal} {Generation}: {The} {Behavior} {Markup} {Language}}, Vol.~\bibinfo{volume}{4133}. \bibinfo{pages}{205--217}.
\newblock
\showISBNx{978-3-540-37593-7}
\urldef\tempurl%
\url{https://doi.org/10.1007/11821830_17}
\showDOI{\tempurl}


\bibitem[Krieglstein et~al\mbox{.}(2022)]%
        {krieglstein_systematic_2022}
\bibfield{author}{\bibinfo{person}{Felix Krieglstein}, \bibinfo{person}{Maik Beege}, \bibinfo{person}{Günter~Daniel Rey}, \bibinfo{person}{Paul Ginns}, \bibinfo{person}{Moritz Krell}, {and} \bibinfo{person}{Sascha Schneider}.} \bibinfo{year}{2022}\natexlab{}.
\newblock \showarticletitle{A {Systematic} {Meta}-analysis of the {Reliability} and {Validity} of {Subjective} {Cognitive} {Load} {Questionnaires} in {Experimental} {Multimedia} {Learning} {Research}}.
\newblock \bibinfo{journal}{\emph{Educational Psychology Review}} \bibinfo{volume}{34}, \bibinfo{number}{4} (\bibinfo{date}{Dec.} \bibinfo{year}{2022}), \bibinfo{pages}{2485--2541}.
\newblock
\showISSN{1573-336X}
\urldef\tempurl%
\url{https://doi.org/10.1007/s10648-022-09683-4}
\showDOI{\tempurl}


\bibitem[Krämer et~al\mbox{.}(2007)]%
        {kramer_effects_2007}
\bibfield{author}{\bibinfo{person}{Nicole Krämer}, \bibinfo{person}{Nina Simons}, {and} \bibinfo{person}{Stefan Kopp}.} \bibinfo{year}{2007}\natexlab{}.
\newblock \showarticletitle{The {Effects} of an {Embodied} {Conversational} {Agent}’s {Nonverbal} {Behavior} on {User}’s {Evaluation} and {Behavioral} {Mimicry}}. In \bibinfo{booktitle}{\emph{Proceedings of the 7th {International} {Conference} on {Intelligent} {Virtual} {Agents}}}. \bibinfo{pages}{238--251}.
\newblock
\showISBNx{978-3-540-74996-7}
\urldef\tempurl%
\url{https://doi.org/10.1007/978-3-540-74997-4_22}
\showDOI{\tempurl}


\bibitem[Krämer et~al\mbox{.}(2003)]%
        {kramer_effects_2003}
\bibfield{author}{\bibinfo{person}{Nicole~C. Krämer}, \bibinfo{person}{Bernd Tietz}, {and} \bibinfo{person}{Gary Bente}.} \bibinfo{year}{2003}\natexlab{}.
\newblock \showarticletitle{Effects of {Embodied} {Interface} {Agents} and {Their} {Gestural} {Activity}}. In \bibinfo{booktitle}{\emph{Intelligent {Virtual} {Agents}}}, \bibfield{editor}{\bibinfo{person}{Thomas Rist}, \bibinfo{person}{Ruth~S. Aylett}, \bibinfo{person}{Daniel Ballin}, {and} \bibinfo{person}{Jeff Rickel}} (Eds.). \bibinfo{publisher}{Springer}, \bibinfo{address}{Berlin, Heidelberg}, \bibinfo{pages}{292--300}.
\newblock
\showISBNx{978-3-540-39396-2}
\urldef\tempurl%
\url{https://doi.org/10.1007/978-3-540-39396-2_49}
\showDOI{\tempurl}


\bibitem[Kucherenko et~al\mbox{.}(2023)]%
        {kucherenko_genea_2023}
\bibfield{author}{\bibinfo{person}{Taras Kucherenko}, \bibinfo{person}{Rajmund Nagy}, \bibinfo{person}{Youngwoo Yoon}, \bibinfo{person}{Jieyeon Woo}, \bibinfo{person}{Teodor Nikolov}, \bibinfo{person}{Mihail Tsakov}, {and} \bibinfo{person}{Gustav~Eje Henter}.} \bibinfo{year}{2023}\natexlab{}.
\newblock \bibinfo{title}{The {GENEA} {Challenge} 2023: {A} large scale evaluation of gesture generation models in monadic and dyadic settings}.
\newblock
\newblock
\urldef\tempurl%
\url{http://arxiv.org/abs/2308.12646}
\showURL{%
\tempurl}
\newblock
\shownote{arXiv:2308.12646 [cs]}.


\bibitem[Kucherenko* et~al\mbox{.}(2024)]%
        {kucherenko2024evaluating}
\bibfield{author}{\bibinfo{person}{Taras Kucherenko*}, \bibinfo{person}{Pieter Wolfert*}, \bibinfo{person}{Youngwoo Yoon*}, \bibinfo{person}{Carla Viegas}, \bibinfo{person}{Teodor Nikolov}, \bibinfo{person}{Mihail Tsakov}, {and} \bibinfo{person}{Gustav~Eje Henter}.} \bibinfo{year}{2024}\natexlab{}.
\newblock \showarticletitle{Evaluating gesture generation in a large-scale open challenge: The GENEA Challenge 2022}.
\newblock \bibinfo{journal}{\emph{ACM Transactions on Graphics}} \bibinfo{volume}{43}, \bibinfo{number}{3} (\bibinfo{year}{2024}), \bibinfo{pages}{1--28}.
\newblock


\bibitem[Kurokawa(1992)]%
        {kurokawa_gesture_1992}
\bibfield{author}{\bibinfo{person}{Takao Kurokawa}.} \bibinfo{year}{1992}\natexlab{}.
\newblock \showarticletitle{Gesture {Coding} and a {Gesture} {Dictionary} for a {Nonverbal} {Interface}}.
\newblock \bibinfo{journal}{\emph{IEICE TRANSACTIONS on Fundamentals of Electronics, Communications and Computer Sciences}} \bibinfo{volume}{E75-A}, \bibinfo{number}{2} (\bibinfo{date}{Feb.} \bibinfo{year}{1992}), \bibinfo{pages}{112--121}.
\newblock
\showISSN{, 0916-8508}
\urldef\tempurl%
\url{https://search.ieice.org/bin/summary.php?id=e75-a_2_112&category=A&year=1992&lang=E&abst=}
\showURL{%
\tempurl}
\newblock
\shownote{Publisher: The Institute of Electronics, Information and Communication Engineers}.


\bibitem[Lin et~al\mbox{.}(2023)]%
        {lin_one-stage_2023}
\bibfield{author}{\bibinfo{person}{Jing Lin}, \bibinfo{person}{Ailing Zeng}, \bibinfo{person}{Haoqian Wang}, \bibinfo{person}{Lei Zhang}, {and} \bibinfo{person}{Yu Li}.} \bibinfo{year}{2023}\natexlab{}.
\newblock \showarticletitle{One-{Stage} {3D} {Whole}-{Body} {Mesh} {Recovery} with {Component} {Aware} {Transformer}}. In \bibinfo{booktitle}{\emph{Proceedings of the {IEEE}/{CVF} {Conference} on {Computer} {Vision} and {Pattern} {Recognition}}}. \bibinfo{pages}{21159--21168}.
\newblock


\bibitem[Liu et~al\mbox{.}(2022)]%
        {liu_beat_2022}
\bibfield{author}{\bibinfo{person}{Haiyang Liu}, \bibinfo{person}{Zihao Zhu}, \bibinfo{person}{Naoya Iwamoto}, \bibinfo{person}{Yichen Peng}, \bibinfo{person}{Zhengqing Li}, \bibinfo{person}{You Zhou}, \bibinfo{person}{Elif Bozkurt}, {and} \bibinfo{person}{Bo Zheng}.} \bibinfo{year}{2022}\natexlab{}.
\newblock \showarticletitle{{BEAT}: {A} {Large}-{Scale} {Semantic} and {Emotional} {Multi}-{Modal} {Dataset} for {Conversational} {Gestures} {Synthesis}}.
\newblock \bibinfo{journal}{\emph{arXiv preprint arXiv:2203.05297}} (\bibinfo{year}{2022}).
\newblock


\bibitem[Liu et~al\mbox{.}(2016)]%
        {liu_two_2016}
\bibfield{author}{\bibinfo{person}{Kris Liu}, \bibinfo{person}{Jackson Tolins}, \bibinfo{person}{Jean~E. Fox~Tree}, \bibinfo{person}{Michael Neff}, {and} \bibinfo{person}{Marilyn~A. Walker}.} \bibinfo{year}{2016}\natexlab{}.
\newblock \showarticletitle{Two {Techniques} for {Assessing} {Virtual} {Agent} {Personality}}.
\newblock \bibinfo{journal}{\emph{IEEE Transactions on Affective Computing}} \bibinfo{volume}{7}, \bibinfo{number}{1} (\bibinfo{date}{Jan.} \bibinfo{year}{2016}), \bibinfo{pages}{94--105}.
\newblock
\showISSN{1949-3045}
\urldef\tempurl%
\url{https://doi.org/10.1109/TAFFC.2015.2435780}
\showDOI{\tempurl}


\bibitem[Liu et~al\mbox{.}(2021)]%
        {liu_speech-based_2021}
\bibfield{author}{\bibinfo{person}{Yu Liu}, \bibinfo{person}{Gelareh Mohammadi}, \bibinfo{person}{Yang Song}, {and} \bibinfo{person}{Wafa Johal}.} \bibinfo{year}{2021}\natexlab{}.
\newblock \showarticletitle{Speech-based {Gesture} {Generation} for {Robots} and {Embodied} {Agents}: {A} {Scoping} {Review}}. In \bibinfo{booktitle}{\emph{Proceedings of the 9th {International} {Conference} on {Human}-{Agent} {Interaction}}} \emph{(\bibinfo{series}{{HAI} '21})}. \bibinfo{publisher}{Association for Computing Machinery}, \bibinfo{address}{New York, NY, USA}, \bibinfo{pages}{31--38}.
\newblock
\showISBNx{978-1-4503-8620-3}
\urldef\tempurl%
\url{https://doi.org/10.1145/3472307.3484167}
\showDOI{\tempurl}


\bibitem[Lund(2007)]%
        {lund_importance_2007}
\bibfield{author}{\bibinfo{person}{Kristine Lund}.} \bibinfo{year}{2007}\natexlab{}.
\newblock \showarticletitle{The importance of gaze and gesture in interactive multimodal explanation}.
\newblock \bibinfo{journal}{\emph{Language Resources and Evaluation}} \bibinfo{volume}{41}, \bibinfo{number}{3-4} (\bibinfo{date}{Dec.} \bibinfo{year}{2007}), \bibinfo{pages}{289--303}.
\newblock
\showISSN{1574-020X, 1572-8412}
\urldef\tempurl%
\url{https://doi.org/10.1007/s10579-007-9058-0}
\showDOI{\tempurl}


\bibitem[Lücking et~al\mbox{.}(2010)]%
        {lucking_bielefeld_2010}
\bibfield{author}{\bibinfo{person}{Andy Lücking}, \bibinfo{person}{Kirsten Bergmann}, \bibinfo{person}{Florian Hahn}, \bibinfo{person}{Stefan Kopp}, {and} \bibinfo{person}{Hannes Rieser}.} \bibinfo{year}{2010}\natexlab{}.
\newblock \showarticletitle{The {Bielefeld} {Speech} and {Gesture} {Alignment} {Corpus} ({SaGA})}.
\newblock \bibinfo{journal}{\emph{LREC 2010 Workshop: Multimodal Corpora–Advances in Capturing, Coding and Analyzing Multimodality}} (\bibinfo{year}{2010}).
\newblock
\urldef\tempurl%
\url{https://pub.uni-bielefeld.de/record/2001935}
\showURL{%
\tempurl}


\bibitem[Maaten and Hinton(2008)]%
        {maaten_visualizing_2008}
\bibfield{author}{\bibinfo{person}{Laurens van~der Maaten} {and} \bibinfo{person}{Geoffrey Hinton}.} \bibinfo{year}{2008}\natexlab{}.
\newblock \showarticletitle{Visualizing {Data} using t-{SNE}}.
\newblock \bibinfo{journal}{\emph{Journal of Machine Learning Research}} \bibinfo{volume}{9}, \bibinfo{number}{86} (\bibinfo{year}{2008}), \bibinfo{pages}{2579--2605}.
\newblock
\showISSN{1533-7928}
\urldef\tempurl%
\url{http://jmlr.org/papers/v9/vandermaaten08a.html}
\showURL{%
\tempurl}


\bibitem[Macedonia(2014)]%
        {macedonia_imitation_2014}
\bibfield{author}{\bibinfo{person}{Manuela Macedonia}.} \bibinfo{year}{2014}\natexlab{}.
\newblock \showarticletitle{Imitation of a {Pedagogical} {Agent}’s {Gestures} {Enhances} {Memory} for {Words} in {Second} {Language}}.
\newblock \bibinfo{journal}{\emph{Science Journal of Education}} \bibinfo{volume}{2}, \bibinfo{number}{5} (\bibinfo{year}{2014}), \bibinfo{pages}{162}.
\newblock
\showISSN{2329-0900}
\urldef\tempurl%
\url{https://doi.org/10.11648/j.sjedu.20140205.15}
\showDOI{\tempurl}


\bibitem[Magid and Pyers(2017)]%
        {magid_i_2017}
\bibfield{author}{\bibinfo{person}{Rachel~W. Magid} {and} \bibinfo{person}{Jennie~E. Pyers}.} \bibinfo{year}{2017}\natexlab{}.
\newblock \showarticletitle{“{I} use it when {I} see it”: {The} role of development and experience in {Deaf} and hearing children’s understanding of iconic gesture}.
\newblock \bibinfo{journal}{\emph{Cognition}}  \bibinfo{volume}{162} (\bibinfo{date}{May} \bibinfo{year}{2017}), \bibinfo{pages}{73--86}.
\newblock
\showISSN{0010-0277}
\urldef\tempurl%
\url{https://doi.org/10.1016/j.cognition.2017.01.015}
\showDOI{\tempurl}


\bibitem[Malkov and Yashunin(2018)]%
        {malkov_efficient_2018}
\bibfield{author}{\bibinfo{person}{Yu~A. Malkov} {and} \bibinfo{person}{D.~A. Yashunin}.} \bibinfo{year}{2018}\natexlab{}.
\newblock \bibinfo{title}{Efficient and robust approximate nearest neighbor search using {Hierarchical} {Navigable} {Small} {World} graphs}.
\newblock
\newblock
\urldef\tempurl%
\url{https://doi.org/10.48550/arXiv.1603.09320}
\showDOI{\tempurl}
\newblock
\shownote{arXiv:1603.09320 [cs]}.


\bibitem[Marcus et~al\mbox{.}(2013)]%
        {marcus_should_2013}
\bibfield{author}{\bibinfo{person}{Nadine Marcus}, \bibinfo{person}{Bejay Cleary}, \bibinfo{person}{Anna Wong}, {and} \bibinfo{person}{Paul Ayres}.} \bibinfo{year}{2013}\natexlab{}.
\newblock \showarticletitle{Should hand actions be observed when learning hand motor skills from instructional animations?}
\newblock \bibinfo{journal}{\emph{Computers in Human Behavior}} \bibinfo{volume}{29}, \bibinfo{number}{6} (\bibinfo{date}{Nov.} \bibinfo{year}{2013}), \bibinfo{pages}{2172--2178}.
\newblock
\showISSN{07475632}
\urldef\tempurl%
\url{https://doi.org/10.1016/j.chb.2013.04.035}
\showDOI{\tempurl}


\bibitem[Matthews-Saugstad et~al\mbox{.}(2017)]%
        {matthews-saugstad_gesturing_2017}
\bibfield{author}{\bibinfo{person}{Krista~M. Matthews-Saugstad}, \bibinfo{person}{Erik~P. Raymakers}, {and} \bibinfo{person}{Damian~G. Kelty-Stephen}.} \bibinfo{year}{2017}\natexlab{}.
\newblock \showarticletitle{Gesturing more diminishes recall of abstract words when gesture is allowed and concrete words when it is taboo}.
\newblock \bibinfo{journal}{\emph{Quarterly Journal of Experimental Psychology}} \bibinfo{volume}{70}, \bibinfo{number}{7} (\bibinfo{date}{July} \bibinfo{year}{2017}), \bibinfo{pages}{1099--1105}.
\newblock
\showISSN{1747-0218}
\urldef\tempurl%
\url{https://doi.org/10.1080/17470218.2016.1263997}
\showDOI{\tempurl}
\newblock
\shownote{Publisher: SAGE Publications}.


\bibitem[Mayer and DaPra(2012)]%
        {mayer_embodiment_2012}
\bibfield{author}{\bibinfo{person}{Richard~E. Mayer} {and} \bibinfo{person}{C.~Scott DaPra}.} \bibinfo{year}{2012}\natexlab{}.
\newblock \showarticletitle{An embodiment effect in computer-based learning with animated pedagogical agents}.
\newblock \bibinfo{journal}{\emph{Journal of Experimental Psychology: Applied}} \bibinfo{volume}{18}, \bibinfo{number}{3} (\bibinfo{year}{2012}), \bibinfo{pages}{239--252}.
\newblock
\showISSN{1939-2192}
\urldef\tempurl%
\url{https://doi.org/10.1037/a0028616}
\showDOI{\tempurl}
\newblock
\shownote{Place: US Publisher: American Psychological Association}.


\bibitem[Mayer and Moreno(1998)]%
        {mayer_split-attention_1998}
\bibfield{author}{\bibinfo{person}{Richard~E. Mayer} {and} \bibinfo{person}{Roxana Moreno}.} \bibinfo{year}{1998}\natexlab{}.
\newblock \showarticletitle{A split-attention effect in multimedia learning: {Evidence} for dual processing systems in working memory.}
\newblock \bibinfo{journal}{\emph{Journal of Educational Psychology}} \bibinfo{volume}{90}, \bibinfo{number}{2} (\bibinfo{date}{June} \bibinfo{year}{1998}), \bibinfo{pages}{312--320}.
\newblock
\showISSN{1939-2176, 0022-0663}
\urldef\tempurl%
\url{https://doi.org/10.1037/0022-0663.90.2.312}
\showDOI{\tempurl}


\bibitem[McInnes et~al\mbox{.}(2020)]%
        {mcinnes_umap_2020}
\bibfield{author}{\bibinfo{person}{Leland McInnes}, \bibinfo{person}{John Healy}, {and} \bibinfo{person}{James Melville}.} \bibinfo{year}{2020}\natexlab{}.
\newblock \bibinfo{title}{{UMAP}: {Uniform} {Manifold} {Approximation} and {Projection} for {Dimension} {Reduction}}.
\newblock
\newblock
\urldef\tempurl%
\url{https://doi.org/10.48550/arXiv.1802.03426}
\showDOI{\tempurl}
\newblock
\shownote{arXiv:1802.03426 [cs, stat]}.


\bibitem[McNeill(1985)]%
        {mcneill_so_1985}
\bibfield{author}{\bibinfo{person}{David McNeill}.} \bibinfo{year}{1985}\natexlab{}.
\newblock \showarticletitle{So you think gestures are nonverbal?}
\newblock \bibinfo{journal}{\emph{Psychological Review}} \bibinfo{volume}{92}, \bibinfo{number}{3} (\bibinfo{year}{1985}), \bibinfo{pages}{350--371}.
\newblock
\showISSN{1939-1471}
\urldef\tempurl%
\url{https://doi.org/10.1037/0033-295X.92.3.350}
\showDOI{\tempurl}
\newblock
\shownote{Place: US Publisher: American Psychological Association}.


\bibitem[Mcneill(1986)]%
        {mcneill_iconic_1986}
\bibfield{author}{\bibinfo{person}{David Mcneill}.} \bibinfo{year}{1986}\natexlab{}.
\newblock \showarticletitle{Iconic gestures of children and adults}.
\newblock  \bibinfo{volume}{62}, \bibinfo{number}{1-2} (\bibinfo{date}{Jan.} \bibinfo{year}{1986}), \bibinfo{pages}{107--128}.
\newblock
\showISSN{1613-3692}
\urldef\tempurl%
\url{https://doi.org/10.1515/semi.1986.62.1-2.107}
\showDOI{\tempurl}
\newblock
\shownote{Publisher: De Gruyter Mouton Section: Semiotica}.


\bibitem[Mousavi et~al\mbox{.}(1995)]%
        {mousavi_reducing_1995}
\bibfield{author}{\bibinfo{person}{Seyed~Yaghoub Mousavi}, \bibinfo{person}{Renae Low}, {and} \bibinfo{person}{John Sweller}.} \bibinfo{year}{1995}\natexlab{}.
\newblock \showarticletitle{Reducing cognitive load by mixing auditory and visual presentation modes.}
\newblock \bibinfo{journal}{\emph{Journal of Educational Psychology}} \bibinfo{volume}{87}, \bibinfo{number}{2} (\bibinfo{date}{June} \bibinfo{year}{1995}), \bibinfo{pages}{319--334}.
\newblock
\showISSN{1939-2176, 0022-0663}
\urldef\tempurl%
\url{https://doi.org/10.1037/0022-0663.87.2.319}
\showDOI{\tempurl}


\bibitem[Neff et~al\mbox{.}(2010)]%
        {neff_evaluating_2010}
\bibfield{author}{\bibinfo{person}{Michael Neff}, \bibinfo{person}{Yingying Wang}, \bibinfo{person}{Rob Abbott}, {and} \bibinfo{person}{Marilyn Walker}.} \bibinfo{year}{2010}\natexlab{}.
\newblock \showarticletitle{Evaluating the effect of gesture and language on personality perception in conversational agents}. In \bibinfo{booktitle}{\emph{Intelligent {Virtual} {Agents}: 10th {International} {Conference}, {IVA} 2010, {Philadelphia}, {PA}, {USA}, {September} 20-22, 2010. {Proceedings} 10}}. \bibinfo{publisher}{Springer}, \bibinfo{pages}{222--235}.
\newblock


\bibitem[Nyatsanga et~al\mbox{.}(2023)]%
        {nyatsanga_comprehensive_2023}
\bibfield{author}{\bibinfo{person}{Simbarashe Nyatsanga}, \bibinfo{person}{Taras Kucherenko}, \bibinfo{person}{Chaitanya Ahuja}, \bibinfo{person}{Gustav~Eje Henter}, {and} \bibinfo{person}{Michael Neff}.} \bibinfo{year}{2023}\natexlab{}.
\newblock \bibinfo{title}{A {Comprehensive} {Review} of {Data}-{Driven} {Co}-{Speech} {Gesture} {Generation}}.
\newblock
\newblock
\urldef\tempurl%
\url{http://arxiv.org/abs/2301.05339}
\showURL{%
\tempurl}
\newblock
\shownote{arXiv:2301.05339 [cs]}.


\bibitem[Oviatt et~al\mbox{.}(2024)]%
        {oviatt_when_2024}
\bibfield{author}{\bibinfo{person}{Sharon Oviatt}, \bibinfo{person}{Rachel Coulston}, {and} \bibinfo{person}{Rebecca Lunsford}.} \bibinfo{year}{2024}\natexlab{}.
\newblock \showarticletitle{When {Do} {We} {Interact} {Multimodally}? {Cognitive} {Load} and {Multimodal} {Communication} {Patterns}}. In \bibinfo{booktitle}{\emph{Proceedings of the 6th {International} {Conference} on {Multimodal} {Interfaces}}}.
\newblock


\bibitem[Paris et~al\mbox{.}(2000)]%
        {paris_linguistic_2000}
\bibfield{author}{\bibinfo{person}{Carol~R. Paris}, \bibinfo{person}{Margaret~H. Thomas}, \bibinfo{person}{Richard~D. Gilson}, {and} \bibinfo{person}{J.~Peter Kincaid}.} \bibinfo{year}{2000}\natexlab{}.
\newblock \showarticletitle{Linguistic {Cues} and {Memory} for {Synthetic} and {Natural} {Speech}}.
\newblock \bibinfo{journal}{\emph{Human Factors}} \bibinfo{volume}{42}, \bibinfo{number}{3} (\bibinfo{date}{Sept.} \bibinfo{year}{2000}), \bibinfo{pages}{421--431}.
\newblock
\showISSN{0018-7208}
\urldef\tempurl%
\url{https://doi.org/10.1518/001872000779698132}
\showDOI{\tempurl}
\newblock
\shownote{Publisher: SAGE Publications Inc}.


\bibitem[Parise et~al\mbox{.}(1999)]%
        {parise_cooperating_1999}
\bibfield{author}{\bibinfo{person}{Salvatore Parise}, \bibinfo{person}{S. Kiesler}, \bibinfo{person}{L. Sproull}, {and} \bibinfo{person}{Keith Waters}.} \bibinfo{year}{1999}\natexlab{}.
\newblock \showarticletitle{Cooperating with life-like interface agents}.
\newblock \bibinfo{journal}{\emph{Computers in Human Behavior}}  \bibinfo{volume}{15} (\bibinfo{year}{1999}), \bibinfo{pages}{123--142}.
\newblock
\urldef\tempurl%
\url{https://doi.org/10.1016/S0747-5632(98)00035-1}
\showDOI{\tempurl}


\bibitem[Perry et~al\mbox{.}(1988)]%
        {perry_transitional_1988}
\bibfield{author}{\bibinfo{person}{Michelle Perry}, \bibinfo{person}{R. Breckinridge~Church}, {and} \bibinfo{person}{Susan Goldin-Meadow}.} \bibinfo{year}{1988}\natexlab{}.
\newblock \showarticletitle{Transitional knowledge in the acquisition of concepts}.
\newblock \bibinfo{journal}{\emph{Cognitive Development}} \bibinfo{volume}{3}, \bibinfo{number}{4} (\bibinfo{date}{Oct.} \bibinfo{year}{1988}), \bibinfo{pages}{359--400}.
\newblock
\showISSN{0885-2014}
\urldef\tempurl%
\url{https://doi.org/10.1016/0885-2014(88)90021-4}
\showDOI{\tempurl}


\bibitem[Ping and Goldin-Meadow(2010)]%
        {ping_gesturing_2010}
\bibfield{author}{\bibinfo{person}{Raedy Ping} {and} \bibinfo{person}{Susan Goldin-Meadow}.} \bibinfo{year}{2010}\natexlab{}.
\newblock \showarticletitle{Gesturing {Saves} {Cognitive} {Resources} {When} {Talking} {About} {Nonpresent} {Objects}}.
\newblock \bibinfo{journal}{\emph{Cognitive Science}} \bibinfo{volume}{34}, \bibinfo{number}{4} (\bibinfo{year}{2010}), \bibinfo{pages}{602--619}.
\newblock
\showISSN{1551-6709}
\urldef\tempurl%
\url{https://doi.org/10.1111/j.1551-6709.2010.01102.x}
\showDOI{\tempurl}
\newblock
\shownote{\_eprint: https://onlinelibrary.wiley.com/doi/pdf/10.1111/j.1551-6709.2010.01102.x}.


\bibitem[Ping and Goldin-Meadow(2008)]%
        {ping_hands_2008}
\bibfield{author}{\bibinfo{person}{Raedy~M. Ping} {and} \bibinfo{person}{Susan Goldin-Meadow}.} \bibinfo{year}{2008}\natexlab{}.
\newblock \showarticletitle{Hands in the air: using ungrounded iconic gestures to teach children conservation of quantity}.
\newblock \bibinfo{journal}{\emph{Developmental Psychology}} \bibinfo{volume}{44}, \bibinfo{number}{5} (\bibinfo{date}{Sept.} \bibinfo{year}{2008}), \bibinfo{pages}{1277--1287}.
\newblock
\showISSN{0012-1649}
\urldef\tempurl%
\url{https://doi.org/10.1037/0012-1649.44.5.1277}
\showDOI{\tempurl}


\bibitem[Radford et~al\mbox{.}(2021)]%
        {radford_learning_2021}
\bibfield{author}{\bibinfo{person}{Alec Radford}, \bibinfo{person}{Jong~Wook Kim}, \bibinfo{person}{Chris Hallacy}, \bibinfo{person}{Aditya Ramesh}, \bibinfo{person}{Gabriel Goh}, \bibinfo{person}{Sandhini Agarwal}, \bibinfo{person}{Girish Sastry}, \bibinfo{person}{Amanda Askell}, \bibinfo{person}{Pamela Mishkin}, \bibinfo{person}{Jack Clark}, \bibinfo{person}{Gretchen Krueger}, {and} \bibinfo{person}{Ilya Sutskever}.} \bibinfo{year}{2021}\natexlab{}.
\newblock \bibinfo{title}{Learning {Transferable} {Visual} {Models} {From} {Natural} {Language} {Supervision}}.
\newblock
\newblock
\urldef\tempurl%
\url{https://doi.org/10.48550/arXiv.2103.00020}
\showDOI{\tempurl}
\newblock
\shownote{arXiv:2103.00020 [cs]}.


\bibitem[Robrecht and Kopp(2023)]%
        {robrecht_snape_2023}
\bibfield{author}{\bibinfo{person}{Amelie Robrecht} {and} \bibinfo{person}{Stefan Kopp}.} \bibinfo{year}{2023}\natexlab{}.
\newblock \showarticletitle{{SNAPE}: {A} {Sequential} {Non}-{Stationary} {Decision} {Process} {Model} for {Adaptive} {Explanation} {Generation}:}. In \bibinfo{booktitle}{\emph{Proceedings of the 15th {International} {Conference} on {Agents} and {Artificial} {Intelligence}}}. \bibinfo{publisher}{SCITEPRESS - Science and Technology Publications}, \bibinfo{address}{Lisbon, Portugal}, \bibinfo{pages}{48--58}.
\newblock
\showISBNx{978-989-758-623-1}
\urldef\tempurl%
\url{https://doi.org/10.5220/0011671300003393}
\showDOI{\tempurl}


\bibitem[Robrecht et~al\mbox{.}(2023)]%
        {robrecht_study_2023}
\bibfield{author}{\bibinfo{person}{Amelie~Sophie Robrecht}, \bibinfo{person}{Markus Rothgänger}, {and} \bibinfo{person}{Stefan Kopp}.} \bibinfo{year}{2023}\natexlab{}.
\newblock \showarticletitle{A {Study} on the {Benefits} and {Drawbacks} {ofAdaptivity} in {AI}-generated {Explanations}}. In \bibinfo{booktitle}{\emph{Proceedings of the 23rd {ACM} {International} {Conference} on {Intelligent} {Virtual} {Agents}}}.
\newblock
\urldef\tempurl%
\url{https://doi.org/10.1145/3570945.3607339}
\showDOI{\tempurl}


\bibitem[Rohlfing et~al\mbox{.}(2021)]%
        {rohlfing_explanation_2021}
\bibfield{author}{\bibinfo{person}{Katharina~J. Rohlfing}, \bibinfo{person}{Philipp Cimiano}, \bibinfo{person}{Ingrid Scharlau}, \bibinfo{person}{Tobias Matzner}, \bibinfo{person}{Heike~M. Buhl}, \bibinfo{person}{Hendrik Buschmeier}, \bibinfo{person}{Elena Esposito}, \bibinfo{person}{Angela Grimminger}, \bibinfo{person}{Barbara Hammer}, \bibinfo{person}{Reinhold Hab-Umbach}, \bibinfo{person}{Ilona Horwath}, \bibinfo{person}{Eyke Hüllermeier}, \bibinfo{person}{Friederike Kern}, \bibinfo{person}{Stefan Kopp}, \bibinfo{person}{Kirsten Thommes}, \bibinfo{person}{Axel~Cyrille Ngonga~Ngomo}, \bibinfo{person}{Carsten Schulte}, \bibinfo{person}{Henning Wachsmuth}, \bibinfo{person}{Petra Wagner}, {and} \bibinfo{person}{Britta Wrede}.} \bibinfo{year}{2021}\natexlab{}.
\newblock \showarticletitle{Explanation as a {Social} {Practice}: {Toward} a {Conceptual} {Framework} for the {Social} {Design} of {AI} {Systems}}.
\newblock \bibinfo{journal}{\emph{IEEE Transactions on Cognitive and Developmental Systems}} \bibinfo{volume}{13}, \bibinfo{number}{3} (\bibinfo{date}{Sept.} \bibinfo{year}{2021}), \bibinfo{pages}{717--728}.
\newblock
\showISSN{23798939}
\urldef\tempurl%
\url{https://doi.org/10.1109/TCDS.2020.3044366}
\showDOI{\tempurl}
\newblock
\shownote{Publisher: Institute of Electrical and Electronics Engineers Inc.}.


\bibitem[Rohlfing et~al\mbox{.}(2006)]%
        {rohlfing_how_2006}
\bibfield{author}{\bibinfo{person}{Katharina~J. Rohlfing}, \bibinfo{person}{Jannik Fritsch}, \bibinfo{person}{Britta Wrede}, {and} \bibinfo{person}{Tanja Jungmann}.} \bibinfo{year}{2006}\natexlab{}.
\newblock \showarticletitle{How can multimodal cues from child-directed interaction reduce learning complexity in robots?}
\newblock \bibinfo{journal}{\emph{Advanced Robotics}} \bibinfo{volume}{20}, \bibinfo{number}{10} (\bibinfo{date}{Jan.} \bibinfo{year}{2006}), \bibinfo{pages}{1183--1199}.
\newblock
\showISSN{0169-1864, 1568-5535}
\urldef\tempurl%
\url{https://doi.org/10.1163/156855306778522532}
\showDOI{\tempurl}


\bibitem[Rosenthal-von~der Pütten and Bergmann(2020)]%
        {rosenthal-von_der_putten_non-verbal_2020}
\bibfield{author}{\bibinfo{person}{Astrid~M. Rosenthal-von~der Pütten} {and} \bibinfo{person}{Kirsten Bergmann}.} \bibinfo{year}{2020}\natexlab{}.
\newblock \showarticletitle{Non-verbal {Enrichment} in {Vocabulary} {Learning} {With} a {Virtual} {Pedagogical} {Agent}}.
\newblock \bibinfo{journal}{\emph{Frontiers in Psychology}}  \bibinfo{volume}{11} (\bibinfo{date}{Nov.} \bibinfo{year}{2020}).
\newblock
\showISSN{1664-1078}
\urldef\tempurl%
\url{https://doi.org/10.3389/fpsyg.2020.533839}
\showDOI{\tempurl}
\newblock
\shownote{Publisher: Frontiers}.


\bibitem[Salem et~al\mbox{.}(2012)]%
        {salem_generation_2012}
\bibfield{author}{\bibinfo{person}{Maha Salem}, \bibinfo{person}{Stefan Kopp}, \bibinfo{person}{Ipke Wachsmuth}, \bibinfo{person}{Katharina Rohlfing}, {and} \bibinfo{person}{Frank Joublin}.} \bibinfo{year}{2012}\natexlab{}.
\newblock \showarticletitle{Generation and {Evaluation} of {Communicative} {Robot} {Gesture}}.
\newblock \bibinfo{journal}{\emph{International Journal of Social Robotics}} \bibinfo{volume}{4}, \bibinfo{number}{2} (\bibinfo{date}{April} \bibinfo{year}{2012}), \bibinfo{pages}{201--217}.
\newblock
\showISSN{1875-4791, 1875-4805}
\urldef\tempurl%
\url{https://doi.org/10.1007/s12369-011-0124-9}
\showDOI{\tempurl}


\bibitem[Sinatra et~al\mbox{.}(2021)]%
        {sinatra_social_2021}
\bibfield{author}{\bibinfo{person}{Anne~M. Sinatra}, \bibinfo{person}{Kimberly~A. Pollard}, \bibinfo{person}{Benjamin~T. Files}, \bibinfo{person}{Ashley~H. Oiknine}, \bibinfo{person}{Mark Ericson}, {and} \bibinfo{person}{Peter Khooshabeh}.} \bibinfo{year}{2021}\natexlab{}.
\newblock \showarticletitle{Social fidelity in virtual agents: {Impacts} on presence and learning}.
\newblock \bibinfo{journal}{\emph{Computers in Human Behavior}}  \bibinfo{volume}{114} (\bibinfo{date}{Jan.} \bibinfo{year}{2021}), \bibinfo{pages}{106562}.
\newblock
\showISSN{0747-5632}
\urldef\tempurl%
\url{https://doi.org/10.1016/j.chb.2020.106562}
\showDOI{\tempurl}


\bibitem[Sweller et~al\mbox{.}(2011)]%
        {sweller_measuring_2011}
\bibfield{author}{\bibinfo{person}{John Sweller}, \bibinfo{person}{Paul Ayres}, {and} \bibinfo{person}{Slava Kalyuga}.} \bibinfo{year}{2011}\natexlab{}.
\newblock \showarticletitle{Measuring {Cognitive} {Load}}.
\newblock In \bibinfo{booktitle}{\emph{Cognitive {Load} {Theory}}}, \bibfield{editor}{\bibinfo{person}{John Sweller}, \bibinfo{person}{Paul Ayres}, {and} \bibinfo{person}{Slava Kalyuga}} (Eds.). \bibinfo{publisher}{Springer}, \bibinfo{address}{New York, NY}, \bibinfo{pages}{71--85}.
\newblock
\showISBNx{978-1-4419-8126-4}
\urldef\tempurl%
\url{https://doi.org/10.1007/978-1-4419-8126-4_6}
\showDOI{\tempurl}


\bibitem[Sweller et~al\mbox{.}(1998)]%
        {sweller_cognitive_1998}
\bibfield{author}{\bibinfo{person}{John Sweller}, \bibinfo{person}{Jeroen J.~G. van Merrienboer}, {and} \bibinfo{person}{Fred G. W.~C. Paas}.} \bibinfo{year}{1998}\natexlab{}.
\newblock \showarticletitle{Cognitive {Architecture} and {Instructional} {Design}}.
\newblock \bibinfo{journal}{\emph{Educational Psychology Review}} \bibinfo{volume}{10}, \bibinfo{number}{3} (\bibinfo{date}{Sept.} \bibinfo{year}{1998}), \bibinfo{pages}{251--296}.
\newblock
\showISSN{1573-336X}
\urldef\tempurl%
\url{https://doi.org/10.1023/A:1022193728205}
\showDOI{\tempurl}


\bibitem[Tindall-Ford et~al\mbox{.}(1997)]%
        {tindall-ford_when_1997}
\bibfield{author}{\bibinfo{person}{Sharon Tindall-Ford}, \bibinfo{person}{Paul Chandler}, {and} \bibinfo{person}{John Sweller}.} \bibinfo{year}{1997}\natexlab{}.
\newblock \showarticletitle{When two sensory modes are better than one.}
\newblock \bibinfo{journal}{\emph{Journal of Experimental Psychology: Applied}} \bibinfo{volume}{3}, \bibinfo{number}{4} (\bibinfo{date}{Dec.} \bibinfo{year}{1997}), \bibinfo{pages}{257--287}.
\newblock
\showISSN{1939-2192, 1076-898X}
\urldef\tempurl%
\url{https://doi.org/10.1037/1076-898X.3.4.257}
\showDOI{\tempurl}


\bibitem[Touvron et~al\mbox{.}(2023)]%
        {touvron_llama_2023}
\bibfield{author}{\bibinfo{person}{Hugo Touvron}, \bibinfo{person}{Louis Martin}, \bibinfo{person}{Kevin Stone}, \bibinfo{person}{Peter Albert}, \bibinfo{person}{Amjad Almahairi}, \bibinfo{person}{Yasmine Babaei}, \bibinfo{person}{Nikolay Bashlykov}, \bibinfo{person}{Soumya Batra}, \bibinfo{person}{Prajjwal Bhargava}, \bibinfo{person}{Shruti Bhosale}, \bibinfo{person}{Dan Bikel}, \bibinfo{person}{Lukas Blecher}, \bibinfo{person}{Cristian~Canton Ferrer}, \bibinfo{person}{Moya Chen}, \bibinfo{person}{Guillem Cucurull}, \bibinfo{person}{David Esiobu}, \bibinfo{person}{Jude Fernandes}, \bibinfo{person}{Jeremy Fu}, \bibinfo{person}{Wenyin Fu}, \bibinfo{person}{Brian Fuller}, \bibinfo{person}{Cynthia Gao}, \bibinfo{person}{Vedanuj Goswami}, \bibinfo{person}{Naman Goyal}, \bibinfo{person}{Anthony Hartshorn}, \bibinfo{person}{Saghar Hosseini}, \bibinfo{person}{Rui Hou}, \bibinfo{person}{Hakan Inan}, \bibinfo{person}{Marcin Kardas}, \bibinfo{person}{Viktor Kerkez}, \bibinfo{person}{Madian Khabsa},
  \bibinfo{person}{Isabel Kloumann}, \bibinfo{person}{Artem Korenev}, \bibinfo{person}{Punit~Singh Koura}, \bibinfo{person}{Marie-Anne Lachaux}, \bibinfo{person}{Thibaut Lavril}, \bibinfo{person}{Jenya Lee}, \bibinfo{person}{Diana Liskovich}, \bibinfo{person}{Yinghai Lu}, \bibinfo{person}{Yuning Mao}, \bibinfo{person}{Xavier Martinet}, \bibinfo{person}{Todor Mihaylov}, \bibinfo{person}{Pushkar Mishra}, \bibinfo{person}{Igor Molybog}, \bibinfo{person}{Yixin Nie}, \bibinfo{person}{Andrew Poulton}, \bibinfo{person}{Jeremy Reizenstein}, \bibinfo{person}{Rashi Rungta}, \bibinfo{person}{Kalyan Saladi}, \bibinfo{person}{Alan Schelten}, \bibinfo{person}{Ruan Silva}, \bibinfo{person}{Eric~Michael Smith}, \bibinfo{person}{Ranjan Subramanian}, \bibinfo{person}{Xiaoqing~Ellen Tan}, \bibinfo{person}{Binh Tang}, \bibinfo{person}{Ross Taylor}, \bibinfo{person}{Adina Williams}, \bibinfo{person}{Jian~Xiang Kuan}, \bibinfo{person}{Puxin Xu}, \bibinfo{person}{Zheng Yan}, \bibinfo{person}{Iliyan Zarov}, \bibinfo{person}{Yuchen
  Zhang}, \bibinfo{person}{Angela Fan}, \bibinfo{person}{Melanie Kambadur}, \bibinfo{person}{Sharan Narang}, \bibinfo{person}{Aurelien Rodriguez}, \bibinfo{person}{Robert Stojnic}, \bibinfo{person}{Sergey Edunov}, {and} \bibinfo{person}{Thomas Scialom}.} \bibinfo{year}{2023}\natexlab{}.
\newblock \bibinfo{title}{Llama 2: {Open} {Foundation} and {Fine}-{Tuned} {Chat} {Models}}.
\newblock
\newblock
\urldef\tempurl%
\url{http://arxiv.org/abs/2307.09288}
\showURL{%
\tempurl}
\newblock
\shownote{arXiv:2307.09288 [cs]}.


\bibitem[van Merriënboer and Sweller(2005)]%
        {van_merrienboer_cognitive_2005}
\bibfield{author}{\bibinfo{person}{Jeroen J.~G. van Merriënboer} {and} \bibinfo{person}{John Sweller}.} \bibinfo{year}{2005}\natexlab{}.
\newblock \showarticletitle{Cognitive {Load} {Theory} and {Complex} {Learning}: {Recent} {Developments} and {Future} {Directions}}.
\newblock \bibinfo{journal}{\emph{Educational Psychology Review}} \bibinfo{volume}{17}, \bibinfo{number}{2} (\bibinfo{date}{June} \bibinfo{year}{2005}), \bibinfo{pages}{147--177}.
\newblock
\showISSN{1573-336X}
\urldef\tempurl%
\url{https://doi.org/10.1007/s10648-005-3951-0}
\showDOI{\tempurl}


\bibitem[Vilhjalmsson et~al\mbox{.}(2007)]%
        {vilhjalmsson_behavior_2007}
\bibfield{author}{\bibinfo{person}{H. Vilhjalmsson}, \bibinfo{person}{N. Cantelmo}, \bibinfo{person}{J. Cassell}, \bibinfo{person}{N.~E. Chafai}, \bibinfo{person}{M. Kipp}, {and} \bibinfo{person}{Stefan Kopp}.} \bibinfo{year}{2007}\natexlab{}.
\newblock \showarticletitle{The {Behavior} {Markup} {Language}: {Recent} {Developments} and {Challenges}}.
\newblock \bibinfo{journal}{\emph{Proc. of Intelligent Virtual Agents (IVA 2007)}}  \bibinfo{volume}{4722} (\bibinfo{year}{2007}).
\newblock
\urldef\tempurl%
\url{https://doi.org/10.1007/978-3-540-74997-4_10}
\showDOI{\tempurl}


\bibitem[Voß and Kopp(2023a)]%
        {vos_aq-gt_2023}
\bibfield{author}{\bibinfo{person}{Hendric Voß} {and} \bibinfo{person}{Stefan Kopp}.} \bibinfo{year}{2023}\natexlab{a}.
\newblock \showarticletitle{{AQ}-{GT}: a {Temporally} {Aligned} and {Quantized} {GRU}-{Transformer} for {Co}-{Speech} {Gesture} {Synthesis}}. In \bibinfo{booktitle}{\emph{Proceedings of the 25th {International} {Conference} on {Multimodal} {Interaction}}} \emph{(\bibinfo{series}{{ICMI} '23})}. \bibinfo{publisher}{Association for Computing Machinery}, \bibinfo{address}{New York, NY, USA}, \bibinfo{pages}{60--69}.
\newblock
\showISBNx{9798400700552}
\urldef\tempurl%
\url{https://doi.org/10.1145/3577190.3614135}
\showDOI{\tempurl}


\bibitem[Voß and Kopp(2023b)]%
        {vos_augmented_2023}
\bibfield{author}{\bibinfo{person}{Hendric Voß} {and} \bibinfo{person}{Stefan Kopp}.} \bibinfo{year}{2023}\natexlab{b}.
\newblock \showarticletitle{Augmented {Co}-{Speech} {Gesture} {Generation}: {Including} {Form} and {Meaning} {Features} to {Guide} {Learning}-{Based} {Gesture} {Synthesis}}. In \bibinfo{booktitle}{\emph{Proceedings of the 23rd {ACM} {International} {Conference} on {Intelligent} {Virtual} {Agents}}}.
\newblock
\urldef\tempurl%
\url{https://doi.org/10.1145/3570945.3607337}
\showDOI{\tempurl}


\bibitem[Wagner et~al\mbox{.}(2014)]%
        {wagner_gesture_2014}
\bibfield{author}{\bibinfo{person}{Petra Wagner}, \bibinfo{person}{Zofia Malisz}, {and} \bibinfo{person}{Stefan Kopp}.} \bibinfo{year}{2014}\natexlab{}.
\newblock \bibinfo{title}{Gesture and speech in interaction: {An} overview}.
\newblock
\newblock
\newblock
\shownote{Pages: 209–232 Publication Title: Speech Communication Volume: 57}.


\bibitem[Wang et~al\mbox{.}(2020)]%
        {wang_minilm_2020}
\bibfield{author}{\bibinfo{person}{Wenhui Wang}, \bibinfo{person}{Furu Wei}, \bibinfo{person}{Li Dong}, \bibinfo{person}{Hangbo Bao}, \bibinfo{person}{Nan Yang}, {and} \bibinfo{person}{Ming Zhou}.} \bibinfo{year}{2020}\natexlab{}.
\newblock \bibinfo{title}{{MiniLM}: {Deep} {Self}-{Attention} {Distillation} for {Task}-{Agnostic} {Compression} of {Pre}-{Trained} {Transformers}}.
\newblock
\newblock
\urldef\tempurl%
\url{https://doi.org/10.48550/arXiv.2002.10957}
\showDOI{\tempurl}
\newblock
\shownote{arXiv:2002.10957 [cs]}.


\bibitem[Wang et~al\mbox{.}(2021)]%
        {wang_understanding_2021}
\bibfield{author}{\bibinfo{person}{Yingfan Wang}, \bibinfo{person}{Haiyang Huang}, \bibinfo{person}{Cynthia Rudin}, {and} \bibinfo{person}{Yaron Shaposhnik}.} \bibinfo{year}{2021}\natexlab{}.
\newblock \showarticletitle{Understanding {How} {Dimension} {Reduction} {Tools} {Work}: {An} {Empirical} {Approach} to {Deciphering} t-{SNE}, {UMAP}, {TriMap}, and {PaCMAP} for {Data} {Visualization}}.
\newblock \bibinfo{journal}{\emph{Journal of Machine Learning Research}} \bibinfo{volume}{22}, \bibinfo{number}{201} (\bibinfo{year}{2021}), \bibinfo{pages}{1--73}.
\newblock
\urldef\tempurl%
\url{http://jmlr.org/papers/v22/20-1061.html}
\showURL{%
\tempurl}


\bibitem[Wickens et~al\mbox{.}(1983)]%
        {wickens_compatibility_1983}
\bibfield{author}{\bibinfo{person}{Christopher~D. Wickens}, \bibinfo{person}{Diane~L. Sandry}, {and} \bibinfo{person}{Michael Vidulich}.} \bibinfo{year}{1983}\natexlab{}.
\newblock \showarticletitle{Compatibility and {Resource} {Competition} between {Modalities} of {Input}, {Central} {Processing}, and {Output}}.
\newblock \bibinfo{journal}{\emph{Human Factors: The Journal of the Human Factors and Ergonomics Society}} \bibinfo{volume}{25}, \bibinfo{number}{2} (\bibinfo{date}{April} \bibinfo{year}{1983}), \bibinfo{pages}{227--248}.
\newblock
\showISSN{0018-7208, 1547-8181}
\urldef\tempurl%
\url{https://doi.org/10.1177/001872088302500209}
\showDOI{\tempurl}


\bibitem[Woods et~al\mbox{.}(2002)]%
        {woods_can_2002}
\bibfield{author}{\bibinfo{person}{David Woods}, \bibinfo{person}{Emily Patterson}, {and} \bibinfo{person}{Emilie Roth}.} \bibinfo{year}{2002}\natexlab{}.
\newblock \showarticletitle{Can {We} {Ever} {Escape} from {Data} {Overload}? {A} {Cognitive} {Systems} {Diagnosis}}.
\newblock \bibinfo{journal}{\emph{Cognition, Technology \& Work}}  \bibinfo{volume}{4} (\bibinfo{date}{April} \bibinfo{year}{2002}), \bibinfo{pages}{22--36}.
\newblock
\urldef\tempurl%
\url{https://doi.org/10.1007/s101110200002}
\showDOI{\tempurl}


\bibitem[Wu et~al\mbox{.}(2014)]%
        {wu_effects_2014}
\bibfield{author}{\bibinfo{person}{Yanxiang Wu}, \bibinfo{person}{Sabarish~V. Babu}, \bibinfo{person}{Rowan Armstrong}, \bibinfo{person}{Jeffrey~W. Bertrand}, \bibinfo{person}{Jun Luo}, \bibinfo{person}{Tania Roy}, \bibinfo{person}{Shaundra~B. Daily}, \bibinfo{person}{Lauren~Cairco Dukes}, \bibinfo{person}{Larry~F. Hodges}, {and} \bibinfo{person}{Tracy Fasolino}.} \bibinfo{year}{2014}\natexlab{}.
\newblock \showarticletitle{Effects of virtual human animation on emotion contagion in simulated inter-personal experiences}.
\newblock \bibinfo{journal}{\emph{IEEE transactions on visualization and computer graphics}} \bibinfo{volume}{20}, \bibinfo{number}{4} (\bibinfo{date}{April} \bibinfo{year}{2014}), \bibinfo{pages}{626--635}.
\newblock
\showISSN{1941-0506}
\urldef\tempurl%
\url{https://doi.org/10.1109/TVCG.2014.19}
\showDOI{\tempurl}


\bibitem[Zhao et~al\mbox{.}(2023)]%
        {zhao_gesture_2023}
\bibfield{author}{\bibinfo{person}{Zeyu Zhao}, \bibinfo{person}{Nan Gao}, \bibinfo{person}{Zhi Zeng}, \bibinfo{person}{Guixuan Zhang}, \bibinfo{person}{Jie Liu}, {and} \bibinfo{person}{Shuwu Zhang}.} \bibinfo{year}{2023}\natexlab{}.
\newblock \showarticletitle{Gesture {Motion} {Graphs} for {Few}-{Shot} {Speech}-{Driven} {Gesture} {Reenactment}}.
\newblock
\urldef\tempurl%
\url{https://openreview.net/forum?id=CMivR3x5fpC}
\showURL{%
\tempurl}


\bibitem[Zhou et~al\mbox{.}(2022)]%
        {zhou_gesturemaster_2022}
\bibfield{author}{\bibinfo{person}{Chi Zhou}, \bibinfo{person}{Tengyue Bian}, {and} \bibinfo{person}{Kang Chen}.} \bibinfo{year}{2022}\natexlab{}.
\newblock \showarticletitle{{GestureMaster}: {Graph}-based {Speech}-driven {Gesture} {Generation}}. In \bibinfo{booktitle}{\emph{{INTERNATIONAL} {CONFERENCE} {ON} {MULTIMODAL} {INTERACTION}}}. \bibinfo{publisher}{ACM}, \bibinfo{address}{Bengaluru India}, \bibinfo{pages}{764--770}.
\newblock
\showISBNx{978-1-4503-9390-4}
\urldef\tempurl%
\url{https://doi.org/10.1145/3536221.3558063}
\showDOI{\tempurl}


\bibitem[Zvaigzne et~al\mbox{.}(2019)]%
        {zvaigzne_how_2019}
\bibfield{author}{\bibinfo{person}{Meghan Zvaigzne}, \bibinfo{person}{Yuriko Oshima-Takane}, {and} \bibinfo{person}{Makiko Hirakawa}.} \bibinfo{year}{2019}\natexlab{}.
\newblock \showarticletitle{How does language proficiency affect children’s iconic gesture use?}
\newblock \bibinfo{journal}{\emph{Applied Psycholinguistics}} \bibinfo{volume}{40}, \bibinfo{number}{2} (\bibinfo{date}{March} \bibinfo{year}{2019}), \bibinfo{pages}{555--583}.
\newblock
\showISSN{0142-7164, 1469-1817}
\urldef\tempurl%
\url{https://doi.org/10.1017/S014271641800070X}
\showDOI{\tempurl}


\end{thebibliography}

\newpage
\appendix

\section{Satisfaction Questionnaire}\label{sec:appendixSQ}
The satisfaction questionnaire is a selection of questions taken from the Artificial Social Agents Questionnaire \cite{fitrianie_artificial-social-agent_2022}, which are translated to German by the authors. Here we will list all original questions used, the test item they are connected to and the translation. In the study the questions were mixed in random order. 
\begin{enumerate}
    \item \textbf{Agent’s Believability}
    \begin{itemize}
        \item [HLB3] \textit{Das Verhalten des Agenten erinnert an menschliches Verhalten.}\\
        The agent’s behavior makes me think of human behavior
        \item [HLB4] \textit{Der Agent verhält sich wie eine echte Person.}\\
        The agent behaves like a real person
        \item [NB2] \textit{Der Agent agiert natürlich.}\\
        The agent acts naturally
    \end{itemize}
    \item \textbf{Performance}
    \begin{itemize}
        \item [PF1] \textit{Der Agent macht seine Aufgabe gut.}\\
        The agent does its task well
        \item [PF2] \textit{Der Agent stört mich nicht.}\\
        The agent does not hinder me.
        \item [PF3] \textit{Mit dem Agenten bin ich in der Lage zu gewinnen.}\\
        I am capable of suceeding with the agent.
    \end{itemize}
    \item \textbf{Likeability}
    \begin{itemize}
        \item [AL2] \textit{Ich mag den Agenten.}\\
        I like the agent
        \item [AL3] \textit{Ich mag den Agenten nicht.}\\
        I dislike the agent
        \item [AL4] \textit{Der Agent ist kooperativ.}\\
        The agent is cooperative
    \end{itemize}
    \item \textbf{User Acceptance of the Agent}
    \begin{itemize}
        \item [UAA1] \textit{Ich würde den Agenten in Zukunft wieder nutzen.}\\
        I will use the agent again in the future
        \item [UAA2] \textit{Ich kann mich den Agenten zukünftig nutzen sehen.}\\
        I can see myself using the agent in the future
        \item [UAA3] \textit{Ich vermeide weitere Interaktionen mit dem Agenten.}\\
        I oppose further interaction with the agent
    \end{itemize}
    \item \textbf{Enjoyability}
    \begin{itemize}
        \item [AE1] \textit{Der Agent ist langweilig.}\\
        The agent is boring
        \item [AE2] \textit{Es ist interessant mit dem Agenten zu interagieren.}\\
        It is interesting to interact with the agent
        \item [AE3] \textit{Ich habe Spaß mit dem Agenten zu interagieren.}\\
        I enjoy interacting with the agent
    \end{itemize}
    \item \textbf{Engagement}
    \begin{itemize}
        \item [UE1] \textit{Während der Interaktion mit dem Agenten war ich konzentriert.}\\
        I was concentrated during the interaction with the agent
        \item [UE2] \textit{Die Interaktion hat meine Aufmerksamkeit erregt.}\\
        The interaction captured my attention
        \item [UE3] \textit{Während der Interaktion war ich aufmerksam.}\\
        I was alert during the interaction with the agent
    \end{itemize}
    \item \textbf{Trust}
    \begin{itemize}
        \item [UT1] \textit{Der Agent gibt gute Hinweise.}\\
        The agent always gives good advice
        \item[UT2] \textit{Der Agent sagt die Wahrheit.}\\
        The agent acts truthfully
        \item[UT3] \textit{Ich kann mich auf den Agenten verlassen.}\\
        I can rely on the agent
    \end{itemize}
    \item \textbf{User-Agent Alliance}
    \begin{itemize}
        \item[UAL2] \textit{Mit dem Agenten zusammenzuarbeiten ist wie ein gemeinsames Projekt.}\\
        Collaborating with the agent is like a joint venture
        \item[UAL4] \textit{Mit dem Agenten kann ich produktiv zusammenarbeiten.}\\
        The agent can collaborate in a productive way
        \item[UAL6]\textit{ Der Agent versteht mich. }\\
        The agent understands me
    \end{itemize}
    \item \textbf{Attentiveness}
    \begin{itemize}
        \item[AA1] \textit{Der Agent ist während der gesamten Interaktion auf mich konzentriert.}\\
        The agent remains focused on me throughout the interaction
        \item[AA2] \textit{Der Agent ist aufmerksam.}\\
        The agent is attentive
        \item[AA3] \textit{Ich bekomme die gesamte Aufmersamkeit des Agenten.}\\
        I receive the agent's full attention throughout the interaction
    \end{itemize}
    \item \textbf{Coherence}
    \begin{itemize}
        \item[AC1] \textit{Das Verhalten des Agenten macht keinen Sinn.}\\
        The agent's behavior does not make sense
        \item[AC3]\textit{ Das Verhalten des Agenten ist inkonsistent.}\\
        The agent is inconsistent
        \item[AC4] \textit{Der Agent wirkt verwirrt.}\\
        The agent appears confused
    \end{itemize}
    \item \textbf{Intentionality}
    \begin{itemize}
        \item[AI1] \textit{Der Agent agiert intentional.}\\
        The agent acts intentionally
        \item[AI3] \textit{Der Agent hat keine Ahnung was er tut.}\\
        The agent has no clue of what it is doing
        \item[AI4] \textit{Der Agent kann eigene Entscheidungen treffen.}\\
        The agent can make its own decision   
    \end{itemize}
    \item \textbf{Social Presence}
    \begin{itemize}
        \item[SP1] \textit{Der Agent hat eine soziale Präsenz.}\\
        The agent has a social presence
        \item[SP2] \textit{Der Agent ist eine soziale Entität.}\\
        The agent is a social entity
        \item[SP3] \textit{Ich habe die gleiche soziale Präsenz wie der Agent}\\
        I have the same social presence as the agent
    \end{itemize}
    \item \textbf{Agent's Emotional Presence}
    \begin{itemize}
        \item[AEP1] \textit{Der Agent ist emotional.}\\ The agent is emotional
        \item[AEP2] \textit{Der Agent hat Emotionen. }\\ The agent experiences emotions
        \item[AEP3] \textit{Der Agent kann keine Emotionen erleben.}\\ The agent cannot experience emotions
    \end{itemize}
    \item \textbf{User's Emotion}
    \begin{itemize}
        \item[UEP1] \textit{Das Auftreten des Agenten hat beeinflusst wie ich mich fühle.}\\
        The agent's attitude influences how I feel
        \item[UEP2] \textit{Ich bin durch die Stimmung des Agenten beeinflusst.}\\
        I am influenced by the agent's moods
        \item[UEP3] \textit{Die Emotionen, die ich während der Interaktion erlebe sind vom Agenten ausgelöst.}\\
        The emotions I feel during the interaction are caused by the agent
    \end{itemize}
\end{enumerate}
 The following questions were added by the authors and focus more on the explanation itself:
 \begin{enumerate}
     \item \textbf{subjective understanding}
     \begin{itemize}
         \item[SU1] \textit{Ich habe die Erklärung gut verstanden.}\\
         I understood the explanation well.
         \item[SU2] \textit{Ich bin nun in der Lage Quarto zu spielen.}\\
         I am now enabled to play Quarto.
         \item[SU3] \textit{Ich habe die Regeln des Spieles noch nicht verstanden.}\\
         I do not understand the rules of the game by now.
     \end{itemize}
     \item \textbf{Agent and Understanding}
     \begin{itemize}
         \item[AU1] \textit{Durch den Agenten habe ich die Erklärung besser verstanden.}\\
         Because of the agent, I understood the explanation better.
         \item[AU2]\textit{Ohne den Agenten hätte ich die Erklärung besser verstehen können.}\\
         It would have been easier to understand without the agent.
         \item[AU3]\textit{Durch den Agenten konnte ich der Erklärung besser folgen.}\\
         Because of the agent, it was easier to follow the explanation.
     \end{itemize}

 \end{enumerate}

\onecolumn
\section{Deep Understanding}\label{sec:appendixU}
The following box plot diagrams show the understanding scores for each of the eight deep understanding items distributed by the four conditions.
\begin{figure*}[htb]
\minipage{0.248\textwidth}
  \includegraphics[width=\linewidth]{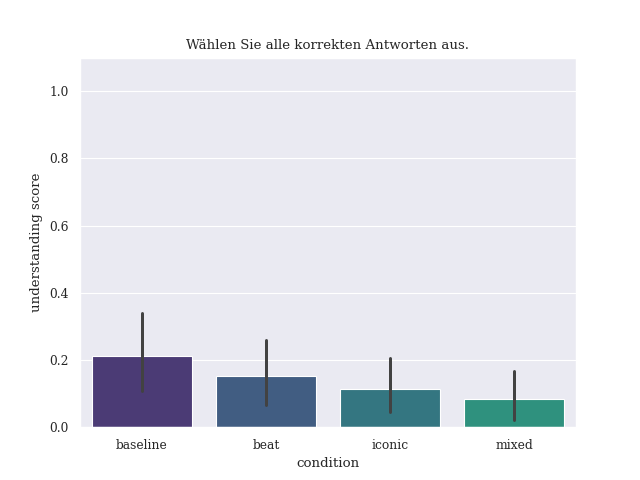}
  \caption{UN30}\label{fig:UN30}
\endminipage
\minipage{0.248\textwidth}
  \includegraphics[width=\linewidth]{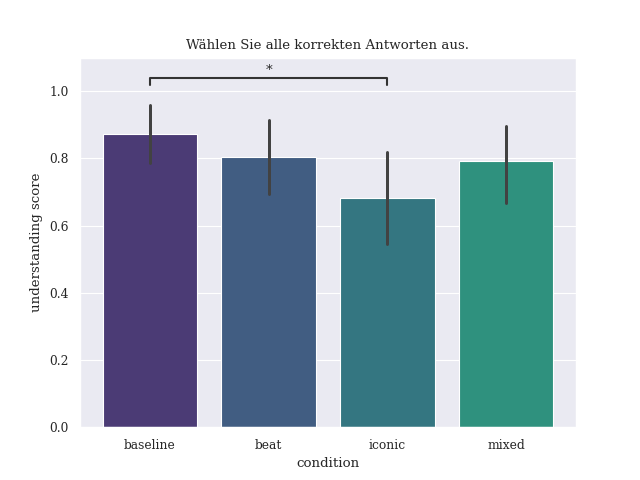}
  \caption{UN31}\label{fig:UN31}
\endminipage
\minipage{0.248\textwidth}%
  \includegraphics[width=\linewidth]{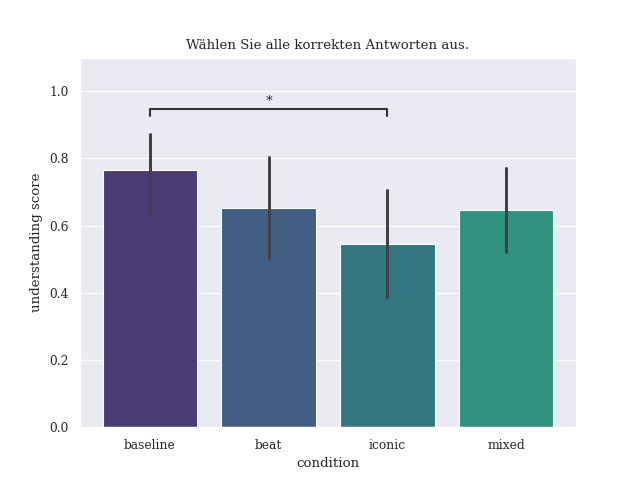}
  \caption{UN32}\label{fig:UN32}
\endminipage
\minipage{0.248\textwidth}%
  \includegraphics[width=\linewidth]{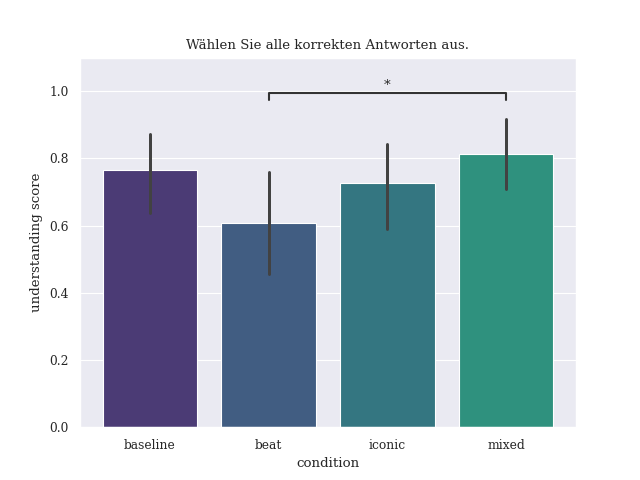}
  \caption{UN33}\label{fig:UN33}
\endminipage
\newline
\minipage{0.248\textwidth}%
  \includegraphics[width=\linewidth]{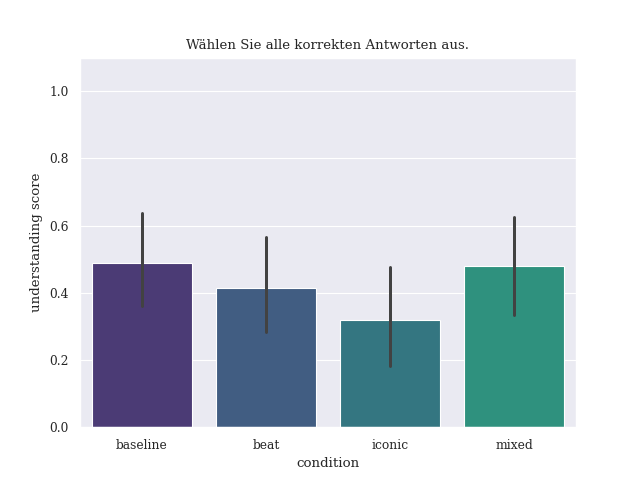}
  \caption{UN34}\label{fig:UN34}
\endminipage
\minipage{0.248\textwidth}%
  \includegraphics[width=\linewidth]{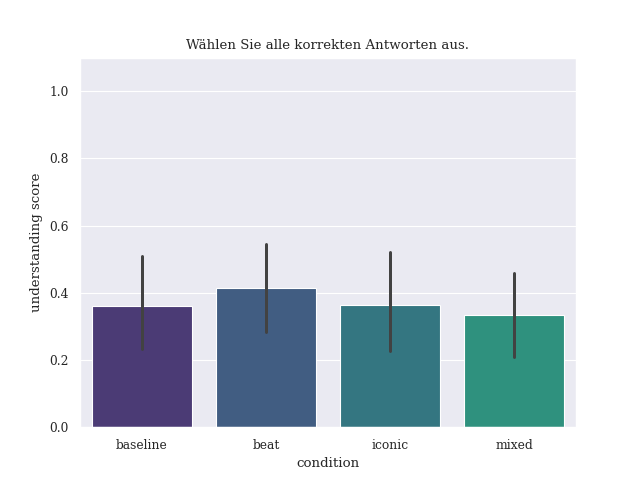}
  \caption{UN35}\label{fig:UN35}
\endminipage
\minipage{0.248\textwidth}%
  \includegraphics[width=\linewidth]{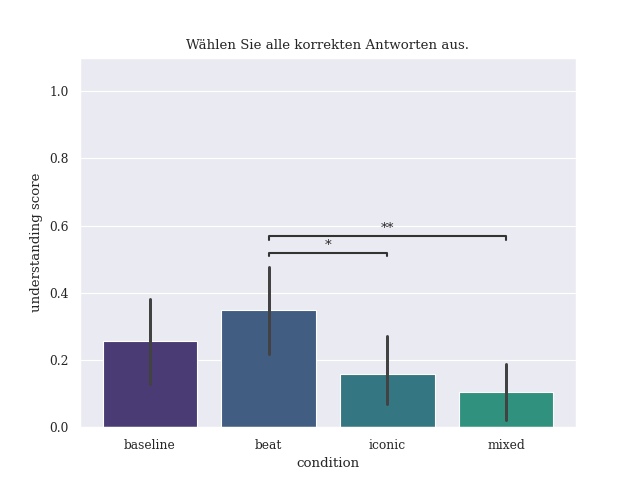}
  \caption{UN36}\label{fig:UN36}
\endminipage
\minipage{0.248\textwidth}%
  \includegraphics[width=\linewidth]{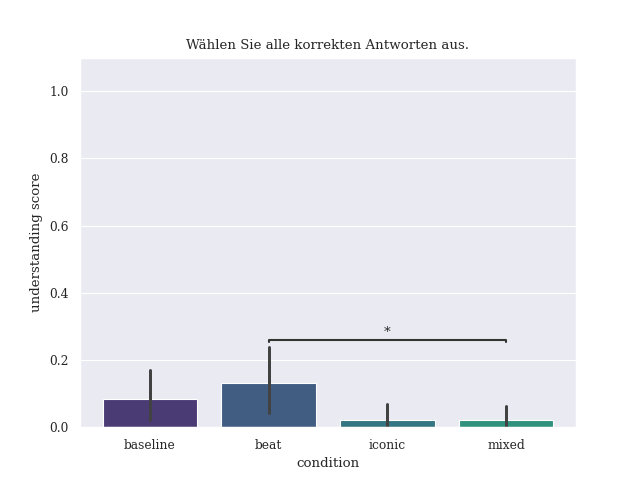}
  \caption{UN37}\label{fig:UN37}
\endminipage
\caption{Comparison between the four conditions for each item in the deep understanding questionnaire}
\label{fig:deep_single}
\end{figure*}

\end{document}